\documentclass{article}
\usepackage{amssymb, amsmath}
\numberwithin{equation}{section}
\newtheorem{theorem}{Theorem}[section]
\newtheorem{proposition}{Proposition}[section]
\newtheorem{lemma}{Lemma}[section]
\newtheorem{remark}{Remark}[section]

\begin{document}
\title{Global Existence of Solutions for the Einstein-Boltzmann System with Cosmological Constant in the Robertson-Walker space-time}
\author{Norbert NOUTCHEGUEME$^{1}$; Etienne TAKOU$^{2}$\\
Department of Mathematics, Faculty of Science, University of
Yaounde 1,\\ PO Box 812, Yaounde, Cameroon \\ {e-mail:
$^{1}$nnoutch@justice.com, $^{1}$nnoutch@uycdc.uninet.cm,
$^{2}$etakou@uycdc.uninet.cm }}
\date{}
\maketitle
\begin{abstract}
We prove a global in time existence theorem for the initial value
problem for the Einstein-Boltzmann system, with cosmological
constant and arbitrarily large initial data, in the spatially
homogeneous case, in a Robertson-Walker space-time.
\end{abstract}
\section{Introduction}
In the mathematical study of General Relativity, one of the main
problems is to establish the existence and to give the properties
of \textbf{global} solutions of the Einstein equations coupled to
various field equations. The knowledge of the global dynamics of
the relativistic kinetic matter is based on such results. In the
case of \textbf{Collisionless}  matter, the phenomena are governed
by the Einstein-Vlasov system in the pure gravitational case, and
by this system coupled to other fields equations, if other fields
than the gravitational field are involved. In the collisionless
case, several authors proved global results, see \cite{a},
\cite{b} for reviews, \cite{c}, \cite{d} and \cite{e} for scalar
matter fields, also see \cite{g}, \cite{h} for the Einstein-Vlasov
system with a cosmological constant. Now in the case of
\textbf{Collisional} matter, the Einstein-Vlasov system is
replaced by the Einstein-Boltzmann system, that seems to be the
best approximation available and that describes the case of
instantaneous, binary and elastic collisions. In contrast with the
abundance of works in the collisionless case, the literature is
very poor in  the collisional case. If, due to its importance in
collisional kinetic theory, several authors studied and proved
global results for the single Boltzmann equation, see \cite{i},
\cite{j} , \cite{k} for the non-relativistic case, and \cite{l},
for the full relativistic case, very few authors studied the
Einstein-Boltzmann system, see \cite{m} for a local existence
theorem. It then seems interesting for us, to extend to the
collisional case, some global results obtained in the
collisionless case. This was certainly the objective of the author
in \cite{n} and \cite{o}, in which he studied the existence of
global solutions of the Einstein-Boltzmann system. Unfortunately,
several points of the work are far from clear; such as, the use of
a formulation which is valid only for the
\textbf{non-relativistic} Boltzmann equation, or, concerning the
Einstein equations, to abandon the evolution equations which are
really relevant, and to concentrate only on the constraint
equations, which, in the homogeneous case studied, reduce as we
will see to a question of choice for the initial data.

In this paper, we study the collisional evolution of a kind of
uncharged \textbf{massive} particles, under the only influence of
their own gravitational field, which is a  function of the
position of the particles.

 The phenomenon is governed, as we said above, by the
coupled Einstein-Boltzmann system we now introduce. The Einstein
equations are the basic equations of the General Relativity. These
equations express the fact that, the gravitational field is
generated by the matter contents, acting as its sources. The
gravitational field is represented, by a second order symmetric
2-tensor of Lorentzian type, called the metric tensor, we denote
by g, whose components $g_{\alpha\beta}$, sometimes called
''gravitational potentials'', are subject to the Einstein
equations, with sources represented by a second order symmetric
2-tensor we denote $T_{\alpha\beta}$, that summarizes all the
matter contents and which is called the stress-matter tensor. Let
us observe that, solving the Einstein equations is determining
both the gravitational field and its sources. In our case, the
only matter contents are the massive particles statistically
described in terms of their \textbf{distribution function},
denoted f, and which is a non-negative real valued function, of
both the position and the momentum of the particles, and that
generates the gravitational field g, through the stress-matter
tensor $T_{\alpha\beta}$. The scalar function f is physically
interpreted as ''probability of the presence density'' of the
particles during their collisional evolution, and is subject to
the Boltzmann equation, defined by a non linear operator Q called
the ''Collision Operator''. In the binary and elastic scheme due
to Lichnerowicz and Chernikov(1940), we adopt, at a given
position, only 2 particles collide, in an instantaneous shock,
without destroying each other, the collision affecting only their
momenta which are not the same, before, and after the shock, only
the sum of the 2 momenta being preserved.

 We then study the coupled Einstein-Boltzmann system in (g,f). The system is
\textbf{coupled} in the sense that f, which is subject to the
Boltzmann equation generates the sources$T_{\alpha\beta}$ of the
Einstein Equations, whereas the metric g, which is subject to the
Einstein equations, is in both the Collision operator, which is
the r.h.s of the Boltzmann equation, and in the Lie derivative of
f with respect to the vectors field tangent to the trajectories of
the particles, and which is the l.h.s of the Boltzmann equation.

We now specify the geometric frame, i.e. the kind of space-time we
are looking for. An important part of general relativity is the
Cosmology, which is the study of the Universe on a large scale. A.
Einstein and W. de Sitter introduced the cosmological models in
1917; A. Friedman and G. Lemaitre introduced the concept of
expanding Universe in 1920. Let us point out the fact that the
Einstein equations are \textbf{overdetermined}, and physically
meaning symmetry assumptions reduce the number of unknowns. A
usual assumption is that the spatial geometry has constant
curvature which is positive, zero or negative, respectively.
Robertson and Walker showed in 1944 that ''exact spherical
symmetry about every point would imply that the universe is
spatially homogeneous'', see \cite{p}, p. 135. We look for a
spatially homogeneous Friedman-Lemaitre-Robertson-Walker
space-time, we will call a ''Robertson-Walker space-time'', which
is, in Cosmology, the basic model for the study of the expanding
Universe. The metric tensor g has only one unknown component we
denote a, which is a \textbf{strictly positive} function called
Cosmological expansion factor; the spatial homogeneity means that
a depends only on the time t and the distribution function f
depends only on the time t and the 4-momentum p of the particles.
The study of the Einstein-Boltzmann system then turns out to the
determination of the couple of scalar function (a, f).

 In the present work, we consider the Einstein Equation with cosmological
constant $\Lambda$. Our motivation is of a physical point of view.
Recent measurements show that the case $\Lambda > 0$ is physically
very interesting in the sense that one can prove, as we will see,
that the expansion of the Universe is accelerating; in
mathematical terms, this means that the mean curvature of the
space-time tends to a constant at late times. For more details on
the cosmological constant, see \cite{q}.

 Let us now sketch the strategy we adopted to prove the global in time existence of a
solution (a, f) of the initial values problem for the
Einstein-Boltzmann system for arbitrarily large initial data
$(a_{0}, f_{0})$ at $t = 0$. In the homogeneous case we consider,
the Einstein Equations are a system of 2 non-linear o.d.e for the
cosmological expansion factor a; the Boltzmann equation is a
non-linear first order p.d.e for the distribution function $f$.

 In a first step, we suppose $a$ given, with the only assumption to be
bounded away from zero, and we give, following Glassey, R.T.,
\cite{r}, the correct formulation of the relativistic Boltzmann
equation in f, on a Robertson-Walker space-time. We then prove
that on any bounded interval $I= [t_{0}, t_{0} + T]$ with $t_{0}
\in \mathbb{R}_{+}$, $T \in \mathbb{R}_{+}^{*}$, the Cauchy
problem for the Boltzmann equation has a unique solution $f \in
C[I; L_{2}^{1}(\mathbb{R}^{3})]$, where
$L_{2}^{1}(\mathbb{R}^{3})$ is a weighted subspace of
$L^{1}(\mathbb{R}^{3})$, whose weight is imposed by the expression
of the sources terms $T_{\alpha\beta}$ of the Einstein equations.
We follow the method developed in \cite{s} that prove a global
existence  of the solution $f \in C[[0, +\infty[;
L^{1}(\mathbb{R}^{3})]$ for the Cauchy problem for the Boltzmann
equation , but here, the norm of $L_{2}^{1}(\mathbb{R}^{3})$ could
allow us to prove the existence theorem only on bounded intervals
$[t_{0}, t_{0} + T]$, and this was enough for the coupling with
the Einstein Equations.

 In a second step, we suppose f given in $C[I; L_{2}^{1}(\mathbb{R}^{3})]$, and we consider the Einstein
equations in a, that split into the constraints equations and the
evolution equation. The constraint equations contain the momentum
constraint which is automatically satisfied in the homogeneous
case we consider, and the Hamiltonian constraint that reduces to a
question of choice for the initial data. The main problem is then
to solve the evolution equation, which is a non-linear second
order o.d.e in a. We set: $e = \frac{1}{a}$, $\theta=
3\frac{\dot{a}}{a}$ (where $\dot{a} = \frac{da}{dt}$), and we
prove that, the Einstein evolution equation in $a$ is equivalent
to the non-linear first order system in (e, $\theta$) defined by
$\dot{e} = - \frac{\theta e}{3}$ and the non-linear first order
Raychaudhuri equation in $\theta$, ($\theta$ is called the
''Hubble variable''), and for which we deduce that there could
exist no global solution in the case $\Lambda < 0$. In the case
$\Lambda > 0$, we deduce by applying standard theorems to the
first order system quoted above, the existence of the solution a
of the evolution Einstein equation, in the space of increasing and
continuous functions on $I = [t_{0}, t_{0} + T]$. We show that $a$
has an exponential growth and the assumption of the first step on
a is then satisfied.

 In a third step, we consider the coupled Einstein-Boltzmann system, and, relying on the above results, we
prove a global in time existence theorem of the solution (a, f) of
the Cauchy problem, by applying the fixed point theorem in an
appropriate function space.

Our method preserves the physical nature of the problem that
imposes to the distribution function f to be a non-negative
function and nowhere, we had to require smallness assumption on
the initial data, which can, consequently, be taken arbitrarily
large.

The paper is organized as follows:\\
In section 2, we introduce the Einstein and Boltzmann equations on
a Robertson-Walker space-time.\\
In section 3, we study the Boltzmann equation in f.\\
In section 4 we study the Einstein equation in a.\\
In section 5, we prove the local existence theorem for the coupled
Einstein-Boltzmann system.\\
In section 6, we prove the global existence theorem for the
coupled Einstein-Boltzmann system.

\section{The Boltzmann equation and the Einstein equations on a Robertson-Walker space-time.}
\subsection{Notations and function spaces}
A greek index varies from 0 to 3 and a Latin index from 1 to 3,
unless otherwise specified. We adopt the Einstein summation
convention $ a_{\alpha}b^{\alpha} = \overset{3}{\underset{\alpha =
0}{\sum}}a_{\alpha}b^{\alpha}$. We consider the flat
Robertson-Walker space-time
denoted ($\mathbb{R}^{4}$, g) where, for \\
$x = (x^{\alpha}) = (x^{0}, x^{i}) \in \mathbb{R}^{4}$, $x^{0} =
t$ denotes the time and $\bar{x} = (x^{i})$ the space. g stands
for the metric tensor with signature (-, +, +, +) that can be
written:
\begin{equation}\label{eq:2.1}
g = - dt^{2} + a^{2}(t)[(dx^{1})^{2} + (dx^{2})^{2} +
(dx^{3})^{2}]
\end{equation}
in which a is a strictly positive function of t, called the
cosmological expansion factor. We consider the
\underline{collisional} evolution of a kind of \textbf{uncharged}
\textbf{massive} relativistic particles in the time oriented
space-time $(\mathbb{R}^{4}, g)$. The particles are statistically
described by their \textbf{distribution function} we denote $f$,
which is a non-negative real-valued function of both the position
$(x^{\alpha})$ and the 4-momentum $p = (p^{\alpha})$ of the
particles, and that coordinatize the tangent bundle
$T(\mathbb{R}^{4})$ i.e:
\begin{equation*}
f: T(\mathbb{R}^{4}) \simeq \mathbb{R}^{4}\times \mathbb{R}^{4}
\rightarrow \mathbb{R}^{+}\,, \quad (x^{\alpha}, p^{\alpha})
\mapsto f(x^{\alpha}, p^{\alpha}) \in \mathbb{R}^{+}
\end{equation*}
For $\bar{p}$ = $(p^{i})$, $\bar{q}$ = $(p^{i})$ $\in
\mathbb{R}^{3}$, we set, as usual:
\begin{equation}\label{eq:2.2}
\bar{p}.\bar{q} = \overset{3}{\underset{i = 1}{\sum}}p^{i}q^{i};
\quad  |\bar{p}| = [\overset{3}{\underset{i =
1}{\sum}}(p^{i})^{2}]^{\frac{1}{2}}
\end{equation}
We suppose the rest mass $m > 0$ of the particles normalized to
unity, i.e\\  we take m = 1. The relativistic particles are then
required to move on the future sheet of the mass-shell whose
equation is
$g(p, p) = -1$. \\
From this, we deduce, using (\ref{eq:2.1}) and (\ref{eq:2.2}):
\begin{equation}\label{eq:2.3}
p^{0} = \sqrt{1 + a^{2}|\bar{p}|^{2}}
\end{equation}
where the choice of $p^{0} > 0$ symbolizes the fact that the
particles eject towards the future . (\ref{eq:2.3}) shows that in
fact, f is defined on the 7-dimensional subbundle of
$T(\mathbb{R}^{4})$ coordinalized by $(x^{\alpha}), (p^{i})$. Now,
we consider the spatially homogeneous case which means that f
depends only on t and $\bar{p} = (p^{i})$. The framework we will
refer to will be the subspace of $L^{1}(\mathbb{R}^{3})$, denote
$L^{1}_{2}(\mathbb{R}^{3})$ and defined by :
\begin{equation} \label{eq:2.4}
L^{1}_{2}(\mathbb{R}^{3}) = \{f \in L^{1}(\mathbb{R}^{3}), \|f\|
:= \int_{\mathbb{R}^{3}}\sqrt{1 + |\bar{p}|^{2}}|f(\bar{p})|d
\bar{p} < + \infty \}
\end{equation}
where $|\bar{p}|$ is given by (\ref{eq:2.2}); $\|.\|$ is a norm on
$L^{1}_{2}(\mathbb{R}^{3})$ and $(L^{1}_{2}(\mathbb{R}^{3}),
\|.\|)$ is a Banach space.\\
Let r be an arbitrary strictly positive real number. We set:
\begin{equation} \label{eq:2.5}
X_{r} = \{ f \in L^{1}_{2}(\mathbb{R}^{3}),
 f \geq 0 \quad a.e, \,  \| f \| \leq r\}
\end{equation}
Endowed with the metric induced by the norm $\|.\|$, $X_{r}$ is a
complete and connected metric subspace of
$(L^{1}_{2}(\mathbb{R}^{3}), \|.\|)$. Let I be a real interval.
Set:
\begin{equation*}
C[I;\, L^{1}_{2}(\mathbb{R}^{3})] = \{ f : I \rightarrow
L^{1}_{2}(\mathbb{R}^{3}), \text{ f continuous and bounded }\}
\end{equation*}
endowed with the norm:
\begin{equation} \label{eq:2.6}
\|| f \|| = \underset{t \in I}{Sup}\|f(t)\|
\end{equation}
$C[I;\, L^{1}_{2}(\mathbb{R}^{3})]$ is a Banach space. $X_{r}$
being defined by (\ref{eq:2.5}). We set:
\begin{equation}\label{eq:2.7}
C[I; X_{r}] = \{ f \in C[I;\, L^{1}_{2}(\mathbb{R}^{3})] , \, f(t)
\in  X_{r}\,,\quad \forall t \in I \}
\end{equation}
Endowed with the metric induced by the norm $\|| . \||$ defined by
(\ref{eq:2.6}), $C[I; X_{r}]$ is a complete metric subspace of
$(C[I; L^{1}_{2}(\mathbb{R}^{3})],\||.\||)$.
\subsection{The Boltzmann equation in ($\mathbb{R}^{4}$, g)}
In its general form, the Boltzmann equation on the curved
space-time ($\mathbb{R}^{4}$, g) can be written:
\begin{equation}\label{eq:2.8}
L_{X}f = Q(f, f)
\end{equation}
where
\begin{equation}\label{eq:2.9}
L_{X}f : = p^{\alpha}\frac{\partial{f}}{\partial{x^{\alpha}}} -
\Gamma^{\alpha}_{\mu\nu}p^{\mu}p^{\nu}\frac{\partial{f}}{\partial{p^{i\alpha}}}
\end{equation}
denotes the Lie derivative of $f$ in the direction of the vector
field X tangent to the trajectories of the particles in
$T\mathbb{R}^{4}$, and whose local coordinates are $(p^{\alpha}, -
\Gamma_{\lambda\mu}^{\alpha}p^{\lambda}p^{\mu})$ in which
$\Gamma_{\lambda\mu}^{\alpha}$ are the Christoffel symbols of g; Q
is a non-linear integral operator called the ''Collision
Operator''. We specify this operator in detail in next section.
Now, since f depends only on t and $(p^{i})$, (\ref{eq:2.8}) can
be written using (\ref{eq:2.9}) :
\begin{equation}\label{eq:2.10}
p^{0}\frac{\partial{f}}{\partial{t}} -
\Gamma^{i}_{\mu\nu}p^{\mu}p^{\nu}\frac{\partial{f}}{\partial{p^{i}}}
= Q(f, f)
\end{equation}
We now express the Christoffel symbols defined by:
\begin{equation}\label{eq:2.11}
\Gamma^{\lambda}_{\alpha\beta}=
\frac{1}{2}g^{\lambda\mu}[\partial_{\alpha}g_{\mu\beta} +
\partial_{\beta}g_{\alpha\mu} - \partial_{\mu}g_{\alpha\beta}]
\end{equation}
in which, the metric g is defined by (\ref{eq:2.1}) and
$g^{\lambda\mu}$ denotes the inverse matrix of $g_{\lambda\mu}$;
(\ref{eq:2.1}) gives:
\begin{equation*}
g^{00} = g_{00} = -1; \quad g_{ii} = a^{2}; \quad g^{ii} = a^{-2};
\quad g_{0i} = g^{0i} = 0; \quad g_{ij} = g^{ij} = 0 \quad for
\quad i \neq j
\end{equation*}
A direct computation, using (\ref{eq:2.11}) then gives, with
$\dot{a} = \frac{da}{dt}$:
\begin{equation}\label{eq:2.12}
\Gamma_{ii}^{0} = \dot{a}a; \quad \Gamma_{i0}^{i} =\Gamma_{0i}^{i}
= \frac{\dot{a}}{a}; \quad \Gamma_{00}^{0} = 0; \quad
\Gamma_{\alpha\beta}^{0} = 0 \quad for \quad \alpha \neq \beta;
\quad \Gamma_{ij}^{k} = 0
\end{equation}
The Boltzmann equation (\ref{eq:2.10}) then writes, using
(\ref{eq:2.12}):
\begin{equation}\label{eq:2.13}
\frac{\partial{f}}{\partial{t}} -
2\frac{\dot{a}}{a}\overset{3}{\underset{i =
1}{\sum}}p^{i}\frac{\partial{f}}{\partial{p^{i}}} =
\frac{1}{p^{0}} Q(f, f)
\end{equation}
in which $p^{0}$ is given by (\ref{eq:2.3}). (\ref{eq:2.13}) is a
non-linear p.d.e in f we study in next section.

\subsection{The Einstein Equations}
We consider the Einstein equations with a \textbf{cosmological
constant $\Lambda$} and that can be written:
\begin{equation}\label{eq:2.14}
R_{\alpha\beta} - \frac{1}{2} R\, g_{\alpha\beta} + \Lambda
g_{\alpha\beta} = 8 \pi G T_{\alpha\beta}
\end{equation}
in which:\\
$R_{\alpha\beta}$ is the Ricci tensor of g, contracted of the
curvature tensor of g;\\
$R = g^{\alpha\beta}R_{\alpha\beta} = R^{\alpha}_{\alpha}$ is the
scalar curvature;\\
$T_{\alpha\beta}$ is the stress-matter tensor that represents the
matter contents, and that is generated by the distribution
function f of the particles by:
\begin{equation}\label{eq:2.15}
T^{\alpha\beta} =
\int_{\mathbb{R}^{3}}\frac{p^{\alpha}p^{\beta}f(t,
\bar{p})|g|^{\frac{1}{2}}}{p^{0}}dp^{1}dp^{2}dp^{3}
\end{equation}
in which $|g|$ is the determinant of g, we have, using
(\ref{eq:2.1}), $|g|^{\frac{1}{2}} = a^{3}$. Recall that f is a
function of t and $\bar{p} = (p^{i})$; then $T^{\alpha\beta}$ is a
function of t.\\
G is the universal gravitational constant. We take G = 1. The
contraction of the Bianchi identities gives the identities
$\nabla_{\alpha}S^{\alpha\beta} = 0$, where $S_{\alpha\beta} =
R_{\alpha\beta} - \frac{1}{2} R\, g_{\alpha\beta}$ is the Einstein
tensor. The Einstein equation (\ref{eq:2.14}) then implies, since
$\nabla g = 0$, that the stress-matter tensor $T^{\alpha\beta}$
must satisfy the four relations $\nabla_{\alpha}T^{\alpha\beta} =
0$ called the conservations laws. But it is proved in \cite{t}
that these laws are satisfied for all solutions f of the Boltzmann
equation. Now if $R_{\quad \alpha,  \mu \beta}^{\lambda}$ are the
components of the curvature tensor of g, we have:
\begin{equation}\label{eq:2.16}
\left\{%
\begin{array}{ll}
    R_{\alpha\beta} &= R_{\quad \alpha,  \lambda \beta}^{\lambda}  \\
    where \\
   R_{\quad \mu,  \alpha \beta}^{\lambda} & =
\partial_{\alpha}{\Gamma^{\lambda}_{\mu\beta}} -
\partial_{\beta}{\Gamma^{\lambda}_{\mu\alpha}} +
\Gamma^{\lambda}_{\nu\alpha}\Gamma^{\nu}_{\mu\beta}-
\Gamma^{\lambda}_{\nu\beta}\Gamma^{\nu}_{\mu\alpha} \\
\end{array}%
\right.
\end{equation}
in which $\Gamma^{\lambda}_{\mu\beta}$ is defined by
(\ref{eq:2.11}). This shows that the Einstein equations
(\ref{eq:2.14}) are a system of non-linear second order p.d.e in
$g_{\alpha\beta}$. In order to have things fresh in mind, when we
will study in details the Einstein equations, we leave the
expression of (\ref{eq:2.14}) in term of a, to section 4 which is
devoted to this study.

\section{Existence Theorem for the Boltzmann Equation}
In this section, we suppose that the cosmological expansion factor
a is given, and we prove an existence theorem for the initial
value problem for the Boltzmann equation (\ref{eq:2.13}), on every
bounded interval $I = [t_{0}, t_{0} + T]$ with $t_{0} \in
\mathbb{R}_{+}$, $T \in \mathbb{R}_{+}$. We begin by specifying
the collision operator Q in (\ref{eq:2.13})
\subsection{The Collision Operator}
In the instanteneous, binary and elastic scheme due to
Lichnerowicz and Chernikov, we consider, at a given position (t,
x), only 2 particles collide instanteneously without destroying
each other, the collision affecting only the momenta of the 2
particles that change after the collision, only the sum of the 2
momenta being preserved, following the scheme:

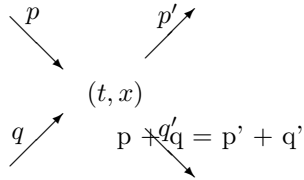
\newcounter{cms}
\setlength{\unitlength}{1mm}

\begin{center}
\begin{picture}(10,20)
\put(11,6.5){\makebox(0,0){$(t,x)$}} \put(-3,-3){\vector(1,1){7}}
\put(-3,17){\vector(1,-1){7}} \put(15,2){\vector(1,-1){6.5}}
\put(15,11.5){\vector(1,1){6.5}} \put(-2,1){\makebox(0,0){$q$}}
\put(0,17){\makebox(0,0){$p$}} \put(18,2){\makebox(0,0){$q'$}}
\put(18,17){\makebox(0,0){$p'$}}
\end{picture}
p + q = p' + q'
\end{center}
The collision operator Q is then defined, using functions f, g on
$\mathbb{R}^{3}$ by:
\begin{equation}\label{eq:3.1}
Q(f, g) = Q^{+}(f, g) - Q^{-}(f, g)
\end{equation}
where
\begin{equation}\label{eq:3.2}
\quad Q^{+}(f, g)(\bar{p}) = \int_{\mathbb{R}^{3}}
\frac{a^{3}d\bar{q}}{q^{0}} \int_{S^{2}}f(\bar{p}')g(\bar{q}'
)A(a, \bar{p}, \bar{q}, \bar{p}', \bar{q}') d\omega
\end{equation}
\begin{equation}\label{eq:3.3}
\quad Q^{-}(f,g)(\bar{p}) =\int_{\mathbb{R}^{3}}.
\frac{a^{3}d\bar{q}}{q^{0}}\int_{S^{2}}f(\bar{p})g(\bar{q})A(a,
\bar{p}, \bar{q}, \bar{p}', \bar{q}')d\omega
\end{equation}
whose elements we now introduce step by step, specifying
properties and hypotheses:
\begin{itemize}
\item[1)] $S^{2}$ is the unit sphere of $\mathbb{R}^{3}$ whose
volume element is denoted $dw$.

\item[2)] A is a non-negative real-valued regular function of all
its arguments, called the \textbf{collision kernel} or the
\textbf{cross-section} of the collisions, on which we require the
following boundedness, symmetry and Lipschitz continuity
assumptions:
\begin{equation}\label{eq:3.4}
0\leq A(a,\, \bar{p},\, \bar{q},\, \bar{p}',\, \bar{q}') \leq
C_{1}
\end{equation}
\begin{equation}\label{eq:3.5}
A(a, \bar{p},\, \bar{q},\, \bar{p}',\, \bar{q}') =A(a, \,
\bar{q},\, \bar{p}, \, \bar{q}',\, \bar{p}')
\end{equation}
\begin{equation}\label{eq:3.6}
A(a, \bar{p},\, \bar{q},\, \bar{p}',\, \bar{q}') =A(a, \,
\bar{p}',\, \bar{q}', \, \bar{p},\, \bar{q})
\end{equation}
\begin{equation}\label{eq:3.7}
|A(a_{1}, \bar{p},\, \bar{q},\, \bar{p}',\, \bar{q}')
 - A(a_{2}, \, \bar{p},\, \bar{q}, \, \bar{p}',\, \bar{q}')|
 \leq \gamma|a_{1} - a_{2}|
\end{equation}
where $C_{1}$ and $\gamma$ are strictly positive constants.

\item[3)]The conservation law $p + q = p' + q'$ splits into:
\begin{equation}\label{eq:3.8}
p^{0} + q^{0} = p'^{0} + q'^{0}
\end{equation}
\begin{equation}\label{eq:3.9}
\bar{p} + \bar{q} = \bar{p}' + \bar{q}'
\end{equation}
and (\ref{eq:3.8}) shows, using (\ref{eq:2.3}), the conservation
of the quantity:
\begin{equation}\label{eq:3.10}
e =  \sqrt{1 + a^{2}\mid\bar{p}\mid^{2}} + \sqrt{1 +
a^{2}\mid\bar{q}\mid^{2}}
\end{equation}
called elementary energy of the unit rest mass partcles; we can
interpret (\ref{eq:3.9}) by setting, following Glassey, R, T., in
\cite{r}:
\begin{equation}\label{eq:3.11}
\begin{cases}
\bar{p}' = \bar{p} + b(\bar{p}, \bar{q}, \omega )\omega\\
\bar{q}' =\bar{q} - b(\bar{p}, \bar{q}, \omega)\omega \, ; \qquad
\omega \in S^{2}
\end{cases}
\end{equation}
in which $b(\bar{p}, \bar{q}, \omega )$ is a real-valued function.
We prove, by a direct calculation, using (\ref{eq:2.3}) to express
$\bar{p}'^{0}$, $\bar{q}'^{0}$ in term of $\bar{p}'$, $\bar{p}'$,
and now (\ref{eq:3.11}) to express $\bar{p}'$, $\bar{q}'$ in term
of $\bar{p}$, $\bar{q}$, that equation (\ref{eq:3.8}) leads to a
quadratic equation in b, that solves to give:
\begin{equation} \label{eq:3.12}
b(\bar{p}, \bar{q}, \omega) = \frac{2\,p^{o}q^{o} e\,a^{2}
\omega.(\hat{\bar{q}} - \hat{\bar{p}})}{e^{2} -
a^{4}(\omega.(\bar{p} + \bar{q}))^{2}}
\end{equation}
in which $\hat{\bar{p}} = \frac{\bar{p}}{p^{0}}$, $\hat{\bar{q}} =
\frac{\bar{q}}{q^{0}}$, and e is given by (\ref{eq:3.10}). Another
direct computation shows, using the classical properties of the
determinants, that the Jacobian of the change of variables
$(\bar{p}, \bar{q}) \rightarrow (\bar{p}', \bar{q}')$ in
$\mathbb{R}^{3}\times\mathbb{R}^{3}$, defined by (\ref{eq:3.11})
is given by:
\begin{equation} \label{eq:3.13}
\frac{\partial(\bar{p}', \bar{q}')}{\partial(\bar{p}, \bar{q})} =
-\frac{p'^{o}q'^{o}}{p^{o}q^{o}}
\end{equation}
\end{itemize}
(\ref{eq:3.13}) shows, using once more (\ref{eq:2.3}) and the
implicit function theorem, that the change of variable
(\ref{eq:3.11}) is invertible and also allows to compute
$\bar{p}$, $\bar{q}$ in term
of $\bar{p}'$, $\bar{q}'$. \\
Finally, formulae (\ref{eq:2.3}) and (\ref{eq:3.11}) show that the
functions to integrate in (\ref{eq:3.2}) and (\ref{eq:3.3})
completely express in terms of $\bar{p}, \bar{q}, \omega$; the
integration with respect to $\bar{q}$ and $\omega$ leave functions
$Q^{+}(f, g)$ and $Q^{-}(f, g)$ of the single variable $\bar{p}$.
In practice, we will consider functions $f$ on  $\mathbb{R} \times
\mathbb{R}^{3}$, that induce, for t fixed in $\mathbb{R}$,
functions f(t) on $\mathbb{R}^{3}$, defined by $f(t)(\bar{p}) =
f(t, \bar{p})$.
\begin{remark}
\begin{itemize}
    \item[1)] Formulae (\ref{eq:3.12}) and (\ref{eq:3.13}) are
    generalizations to the case of the Robertson-Walker
    space-time, of analogous formulae established by Glassey, R.T,
    in \cite{r}, in the case of the Minkowski space-time, to which
    the Robertson-Walker space-time reduces, when we take $a(t) =
    1$ in the metric (\ref{eq:2.1}).
    \item[2)] $A = k e^{- a^{2} - |\bar{p}|^{2} - |\bar{q}|^{2} - |\bar{p}'|^{2} -
    |\bar{q}'|^{2}}$, $k > 0$, is a simple example of functions
    satisfying assumptions (\ref{eq:3.4}), (\ref{eq:3.5}),
    (\ref{eq:3.6}) and (\ref{eq:3.7}).
\end{itemize}
\end{remark}
\subsection{Resolution of the Boltzmann equation}
We consider the Boltzmann equation on $[t_{0}, t_{0} + T]$ with
$t_{0} \in \mathbb{R}_{+}$, $T \in \mathbb{R}_{+}^{*}$ and a is
supposed to be given and defined on $[t_{0}, t_{0} + T]$.\\
The Boltzmann equation (\ref{eq:2.13}) is a first order p.d.e and
its resolution is equivalent to the resolution of the associated
characteristic system, which can be written, taking t as
parameter:
\begin{equation}\label{eq:3.14}
\frac{dp^{i}}{dt} = -2\,\,\frac{\dot{a}}{a}; \qquad
 \frac{df}{dt} =
\frac{1}{p^{0}}Q(f, f)
\end{equation}
We solve the initial value problem on $I = [t_{0}, t_{0} + T]$
with initial data:
\begin{equation}\label{eq:3.15}
p^{i}(t_{0}) = y^{i}; \qquad
 f(t_{0}) = f_{t_{0}}
\end{equation}
The equation in $\bar{p} = (p^{i})$ solve directly to give,
setting $y = (y^{i}) \in \mathbb{R}^{3}$;
\begin{equation}\label{eq:3.16}
\bar{p}(t_{0} + t, y) = \frac{a^{2}(t_{0})}{a^{2}(t_{0} + t)}y,
\quad t\in [0, T]
\end{equation}
The initial value problem for f is equivalent to the following
integral equation in f, in which $\bar{p}$ stands  this time for
any independent variable in $\mathbb{R}^{3}$:
\begin{equation}\label{eq:3.17}
f(t_{0} + t, \bar{p}) = f_{t_{0}}(\bar{p}) + \int_{t_{0}}^{t_{0} +
t}\frac{1}{p^{0}}Q(f, f)(s, \bar{p})ds \quad t \in [0, T]
\end{equation}
Finally, solving the Boltzmann equation (\ref{eq:2.13}) is
equivalent  to solving the integral equation (\ref{eq:3.17}). We
prove:
\begin{theorem}
Let a be a strictly positive continuous function such that $a(t)
\geq \frac{3}{2}$ whenever a is defined .\\
Let $f_{t_{0}} \in L_{2}^{1}(\mathbb{R}^{3})$, $f_{t_{0}} \geq 0$,
a.e,  $r \in \mathbb{R}^{*}_{0}$ such that $r > \|f_{t_{0}}\|$.
Then, the initial value problem for the Boltzmann equation on
$[t_{0}, t_{0} + T]$, with initial data $f_{t_{0}}$, has a unique
solution $f \in C[[t_{0}, t_{0} + T]; X_{r}]$. Moreover, f
satisfies the estimation:
\begin{equation}\label{eq:3.18}
\underset{t \in [t_{0}, t_{0} + T]}{Sup}\|f(t)\| \leq
\|f_{t_{0}}\|
\end{equation}
\end{theorem}
Theorem 3.1 is a direct consequence of the following result:
\begin{proposition}
Assume hypotheses of theorem 3.1 on: $a$, $f_{t_{0}}$ and r.
\begin{itemize}
    \item[1)] There exists an integer $n_{0}(r)$ such that, for
    every integer $n \geq n_{0}(r)$ and for every $v \in X_{r}$,
    the equation
    \begin{equation}\label{eq:3.19}
\sqrt{n} u - \frac{1}{p^{0}}Q(u, u) = v
\end{equation}
has a unique solution  $u_{n} \in X_{r}$
    \item[2)] Let  $n \in \mathbb{N}$, $n \geq n_{0}(r)$
    \begin{itemize}
        \item[i)] For every $u \in X_{r}$ , define R(n, Q)u to be
        the unique element of $X_{r}$ such that:
        \begin{equation}\label{eq:3.20}
 \sqrt{n}R(n, Q)u - \frac{1}{p^{0}}Q\left[R(n, Q)u, R(n, Q)u\right] = u
\end{equation}
        \item[ii)] Define operator $Q_{n}$ on $X_{r}$ by:
        \begin{equation}\label{eq:3.21}
 Q_{n}(u, u) = n\sqrt{n}R(n, Q)u - n u
\end{equation}
Then
\begin{itemize}
    \item[a)] The integral equation
    \begin{equation}\label{eq:3.22}
f(t_{0} + t, \bar{p}) = f_{t_{0}}(\bar{p}) + \int_{t_{0}}^{t_{0} +
t}Q_{n}(f, f)(s, \bar{p})ds  \qquad t\in[0, T]
\end{equation}
has a unique solution $f_{n} \in C[[t_{0}, t_{0} + T]; X_{r}]$.\\
Moreover, $f_{n}$ satisfies the estimation:
\begin{equation}\label{eq:3.23}
\underset{t \in [t_{0}, t_{0} + T]}{Sup}\|f_{n}(t)\| \leq
\|f_{t_{0}}\|
\end{equation}
    \item[b)]The sequence $(f_{n})$ converges in $C[[t_{0}, t_{0} + T];
    X_{r}]$to an element $f \in C[[t_{0}, t_{0} + T]; X_{r}]$, which
    is the unique solution of the integral equation (\ref{eq:3.17}).
    The solution f satisfies  the estimation (\ref{eq:3.18}).
\end{itemize}
\end{itemize}
\end{itemize}
\end{proposition}
\textbf{Proof of the proposition 3.1}\\
The proof follows the same lines as the proof of theorem 4.1 in
\cite{s}. We will emphasize only on points where differences arise
with the present case and we show how we proceed in such cases.\\
\textbf{Proof of point 1) of prop 3.1}\\
We use:
\begin{lemma}
Let $f, g \in L^{1}_{2}(\mathbb{R}^{3})$. then
$\frac{1}{p^{0}}Q^{+}(f, g), \, \, \frac{1}{p^{0}}Q^{-}(f, g) \in
L^{1}_{2}(\mathbb{R}^{3})$ and
\begin{equation}\label{eq:3.24}
\left \| \frac{1}{p^{0}}Q^{+}(f, g) \right \| \leq C(t)\parallel f
\parallel
\parallel g \parallel, \quad \left \| \frac{1}{p^{0}}Q^{-}(f, g) \right \| \leq C(t)\parallel f
\parallel
\parallel g \parallel
\end{equation}

\begin{equation}\label{eq:3.25}
\begin{cases}
 \left \| \frac{1}{p^{0}}Q^{+}(f, f) -
\frac{1}{p^{0}}Q^{+}(g, g)\right \| \leq C(t)(\parallel f
\parallel + \parallel g
\parallel)\parallel f - g \parallel\\
\\
\left \| \frac{1}{p^{0}}Q^{-}(f, f) - \frac{1}{p^{0}}Q^{-}(g,
g)\right \| \leq C(t)(\parallel f \parallel + \parallel g
\parallel)\parallel f - g \parallel
\end{cases}
\end{equation}
\begin{equation} \label{eq:3.26}
\left \| \frac{1}{p^{0}}Q(f, f) - \frac{1}{p^{0}}Q(g, g)\right \|
\leq C(t)(\parallel f \parallel + \parallel\ g
\parallel)
\parallel f - g \parallel
\end{equation}
where
\begin{equation}\label{eq:3.27}
C(t)= 32 \pi C_{1} a^{3}(t)
\end{equation}
\end{lemma}
\textbf{proof of lemma 3.1}\\
We deduce from (\ref{eq:3.8}) and $a > 1$ that:
\begin{equation*}
\sqrt{1 + |\bar{p}|^{2}} \leq \sqrt{1 + \frac{1}{a^{2}}}p^{0} \leq
\sqrt{1 + \frac{1}{a^{2}}}(p^{0} + q^{0}) = \sqrt{1 +
\frac{1}{a^{2}}}(p'^{0} + q'^{0}) \leq  2(p'^{0} + q'^{0})
\end{equation*}
Expression (\ref{eq:3.2}) of $Q^{+}(f, g)$ then gives, using
(\ref{eq:3.4}):
\begin{equation*}
\left\|\frac{1}{p^{0}}Q^{+}(f, g)\right\| =
\int_{\mathbb{R}^{3}}\left|\frac{\sqrt{1 +
|\bar{p}|^{2}}}{p^{0}}Q^{+}(f, g)\right|d\bar{p} \leq
2a^{3}(t)C_{1}\int_{\mathbb{R}^{3}}\int_{\mathbb{R}^{3}}\int_{S^{2}}\frac{p'^{0}
+
q'^{0}}{p^{0}q^{0}}|f(\bar{p}')||g(\bar{q}')|d\bar{p}d\bar{q}d\omega
\end{equation*}
We then deduce, using the change of variables (\ref{eq:3.11}) and
(\ref{eq:3.13}) that gives $d\bar{p}d\bar{q} \, = \,
\frac{p^{0}q^{0}}{p'^{0}q'^{0}}d\bar{p}'d\bar{q}' $ and the fact
that, by (\ref{eq:2.3}) $\frac{p'^{0} + q'^{0}}{p'^{0}q'^{0}}  =
\frac{1}{q'^{0}} + \frac{1}{q'^{0}} \leq 2$
\begin{equation*}
\left\|\frac{1}{p^{0}}Q^{+}(f, g)\right\| \leq 8\pi
a^{3}(t)C_{1}\int_{\mathbb{R}^{3}}\int_{\mathbb{R}^{3}}
|f(\bar{p}')||g(\bar{q}')|d\bar{p}'d\bar{q}' \leq C(t)\| f \|g\|
\end{equation*}
The estimation of $\left\|\frac{1}{p^{0}}Q^{-}(f, g)\right\|$
follows the same way without change of variables and
(\ref{eq:3.24}) follows. The inequalities (\ref{eq:3.25}) are
consequences of (\ref{eq:3.24}) and the bilinearity of $Q^{+}$ and
$Q^{-}$, that allows us to write, P
standing for $\frac{1}{p^{0}}Q^{+}$ or $\frac{1}{p^{0}}Q^{-}$:\\
$P(f, f) - P(g, g) = P(f, f- g) + P(f - g, g)$.\\
Finally, (\ref{eq:3.26}) is a consequence of (\ref{eq:3.25}) and
$Q = Q^{+} - Q^{-}$. This completes the proof of the lemma 3.1.\\
Now the continuous and strictly positive function $t \rightarrow
a^{3}(t)$ is bounded from above, on the line segment $[t_{0},
t_{0} + T]$, and  C(t) given by (\ref{eq:3.27}) is bounded from
above by a constant $C(t_{0}, T) > 0$. Hence, if we replace C(t)
by $C(t_{0}, T)$ in the inequalities in lemma 3.1, we obtain the
same inequalities with an absolute constant than the inequalities
in proposition 3.1 in \cite{s}. The proof of the point 1) of prop
3.1 is then exactly
the same as the proof of prop 3.2 in \cite{s}. $\blacksquare$\\
\textbf{Proof of the point 2) of prop 3.1}\\
We use, $n_{0}(r)$ being the integer introduced in point 1:
\begin{lemma}
We have, for every integer $n \geq n_{0}(r)$ and for every $u \in
X_{r}$
\begin{equation}\label{eq:3.28}
\parallel \sqrt{n}R(n, Q)u\parallel = \parallel u \parallel
\end{equation}
\end{lemma}
\textbf{Proof of lemma 3.2}\\
(\ref{eq:3.28}) is a consequence of:
\begin{equation} \label{eq:3.29}
\int_{\mathbb{R}^{3}}Q(f, f)(\bar{p})d \bar{p} \, =\, 0\,, \qquad
\forall f \in L^{1}_{2}(\mathbb{R}^{3})
\end{equation}
we now establish. It is here that assumption (\ref{eq:3.5}),
(\ref{eq:3.6}) on the collision kernel A are required. Define
operator $Q^{*}$, for $f, g \in L^{1}_{2}(\mathbb{R}^{3})$ by:
\begin{equation*}
\begin{aligned}
Q^{*}(f, g)(\bar{p}) &=
\frac{1}{2}\int_{\mathbb{R}^{3}}\int_{S^{2}}\frac{a^{3}A(a,
\bar{p}, \bar{q}, \bar{p}',
\bar{q}')}{q^{0}}[f(\bar{p}')g(\bar{q}') +
f(\bar{q}')g(\bar{p}')\\
&-f(\bar{p})g(\bar{q}) - f(\bar{q})g(\bar{p})]d\bar{q}d\omega
\end{aligned} \tag{a}
\end{equation*}
Clearly, $Q^{*}(f, f) = Q(f, f)$, where Q is the collision
operator defined by (\ref{eq:3.1}), (\ref{eq:3.2}),
(\ref{eq:3.3}). Now let $\Phi$ be a regular function on
$\mathbb{R}^{3}$, such that:
\begin{equation*}
\Phi (\bar{p}) + \Phi (\bar{q}) = \Phi (\bar{p}') + \Phi
(\bar{q}') \tag{b}
\end{equation*}
Multiplying (a) by $\frac{1}{p^{0}}\Phi (\bar{p})$ and integrating
on $\mathbb{R}^{3}$ yields:
\begin{equation*}
\begin{aligned}
\int_{\mathbb{R}^{3}}\frac{1}{p^{0}}Q^{*}(f, g)(\bar{p})\Phi
(\bar{p})d\bar{p} &= \frac{a^{3}}{2}\int_{\mathbb{R}^{3}\times
\mathbb{R}^{3}\times S^{2}}\frac{A(a, \bar{p}, \bar{q}, \bar{p}',
\bar{q}')}{p^{0}q^{0}}[f(\bar{p}')g(\bar{q}') +
f(\bar{q}')g(\bar{p}') \\
&- f(\bar{p})g(\bar{q}) - f(\bar{q})g(\bar{p})]\Phi
(\bar{p})d\bar{p}d\bar{q}d\omega
\end{aligned} \tag{c}
\end{equation*}
Let us compute the integral of the r.h.s of (c) using the change
of variables $(\bar{p}, \bar{q}) \mapsto (\bar{q}, \bar{p})$;
formulae (\ref{eq:3.12}) shows, since $p^{0}q^{0}$ remains
unchanged, that $b(\bar{p}, \bar{q}, \omega)$ change to
 $-b(\bar{p}, \bar{q}, \omega)$ and formulae (\ref{eq:3.11}) then
 show that $(\bar{p}', \bar{q}')$ change to $(\bar{q}',
 \bar{p}')$; next assumption (\ref{eq:3.5}) on $A$ show that $A(a, \bar{p}, \bar{q}, \bar{p}',
 \bar{q}')$ remains unchanged. We then have, since the Jacobian of
 the change of variables $(\bar{p}, \bar{q}) \rightarrow (\bar{q},
\bar{p})$ is 1
\begin{equation*}
\begin{aligned} \int_{\mathbb{R}^{3}}\frac{1}{p^{0}}Q^{*}(f, g)\Phi
(\bar{p})d\bar{p} &= \frac{a^{3}}{2}\int_{\mathbb{R}^{3}\times
\mathbb{R}^{3}\times S^{2}}\frac{A(a, \bar{p}, \bar{q}, \bar{p}',
\bar{q}')}{p^{0}q^{0}}[f(\bar{p}')g(\bar{q}') +
f(\bar{q}')g(\bar{p}') \\
&- f(\bar{p})g(\bar{q}) - f(\bar{q})g(\bar{p})]\Phi
(\bar{q})d\bar{p}d\bar{q}d\omega
\end{aligned} \tag{d}
\end{equation*}
The sum of (c) and  (d) gives:
\begin{equation*}
\begin{aligned}
\int_{\mathbb{R}^{3}}\frac{1}{p^{0}}Q^{*}(f, g)\Phi
(\bar{p})d\bar{p} &= \frac{a^{3}}{4}\int_{\mathbb{R}^{3}\times
\mathbb{R}^{3} \times S^{2}}\frac{S(a, \bar{p}, \bar{q}, \bar{p}',
\bar{q}')}{p^{0}q^{0}}[f(\bar{p}')g(\bar{q}') +
f(\bar{q}')g(\bar{p}') \\
&- f(\bar{p})g(\bar{q}) - f(\bar{q})g(\bar{p})]
d\bar{p}d\bar{q}d\omega
\end{aligned} \tag{e}
\end{equation*}
where
\begin{equation*}
S(a, \bar{p}, \bar{q}, \bar{p}', \bar{q}') = A(a, \bar{p},
\bar{q}, \bar{p}', \bar{q}')[ \Phi (\bar{p})+ \Phi (\bar{q})]
\tag{f}
\end{equation*}
we can write using, (e)
\begin{equation*}
\int_{\mathbb{R}^{3}}\frac{1}{p^{0}}Q^{*}(f, g)(\bar{p})\Phi
(\bar{p})d\bar{p} = I_{0} - J_{0} \tag{g}
\end{equation*}
where
\begin{equation*}
\begin{cases}
I_{0} = \frac{a^{3}}{4}\int_{\mathbb{R}^{3}}
\int_{\mathbb{R}^{3}}\int_{S^{2}}\frac{S(a, \bar{p}, \bar{q},
\bar{p}', \bar{q}')}{p^{0}q^{0}}[f(\bar{p}')g(\bar{q}') +
f(\bar{q}')g(\bar{p}')]d\bar{p}d\bar{q}d\omega\\
\\
J_{0} = \frac{a^{3}}{4}\int_{\mathbb{R}^{3}}
\int_{\mathbb{R}^{3}}\int_{S^{2}}\frac{S(a, \bar{p}, \bar{q},
\bar{p}', \bar{q}')}{p^{0}q^{0}}[f(\bar{p})g(\bar{q}) +
f(\bar{q})g(\bar{p})]d\bar{p}d\bar{q}d\omega
\end{cases}
\end{equation*}
Let us make in $I_{0}$ the change of variable $(\bar{p}, \bar{q})
\rightarrow (\bar{p}', \bar{q}')$ defined by (\ref{eq:3.11}) and
whose Jacobian (\ref{eq:3.13}) gives  $d\bar{p}d\bar{q} =
\frac{p^{0}q^{0}}{p'^{0}q'^{0}}d\bar{p}'d\bar{q}'$
 ; we have, since $(\bar{p}, \bar{q})$ is expressible in term of  $(\bar{p}', \bar{q}')$ by virtue of (\ref{eq:3.13}):
\begin{equation*}
I_{0} = \frac{a^{3}}{2}\int_{S^{2}}d\omega
\int_{\mathbb{R}^{3}\times \mathbb{R}^{3}}\frac{S(a, \bar{p},
\bar{q}, \bar{p}', \bar{q}')}{p'^{0}q'^{0}}[f(\bar{p}')g(\bar{q}')
+ f(\bar{q}')g(\bar{p}')]d\bar{p}'d\bar{q}' \tag{h}
\end{equation*}
Now, assumption (\ref{eq:3.6}) on $A$ and assumption (b) on $\Phi$
imply that  $S$ defined by  (f) satisfies the relation: $S(a,
\bar{p}, \bar{q}, \bar{p}', \bar{q}') = S(a, \bar{p}', \bar{q}',
\bar{p}, \bar{q})$. Replacing in $I_{0}$  $S(a, \bar{p}, \bar{q},
\bar{p}', \bar{q}')$ by $S(a, \bar{p}', \bar{q}', \bar{p},
\bar{q})$ , we deduce, from the expression (h) of $I_{0}$ that
$I_{0} = J_{0}$. then (g) implies:
\begin{equation*}
\int_{\mathbb{R}^{3}}\frac{1}{p^{0}}Q^{*}(f, g)\Phi
(\bar{p})d\bar{p} = 0 \tag{i}
\end{equation*}

Now the function $\Phi (\bar{p}) = p^{0} = \sqrt{1 +
a^{2}|\bar{p}|^{2}}$ satisfies hypothesis (b) as a consequence of
the conservation law (\ref{eq:3.8}). The relation (\ref{eq:3.29})
then follows from the above choice of $\Phi(\bar{p})$, (i) and the
relation $Q^{*}(f, f) = Q(f,
f)$.\\
Now let us prove (\ref{eq:3.28}).\\
We have, multiplying equation (\ref{eq:3.20}) by $p^{0} = \sqrt{1
+ a^{2}|\bar{p}|^{2}}$, integrating over $\mathbb{R}^{3}$, and
using (\ref{eq:3.29}):
\begin{equation}\label{eq:3.30}
\sqrt{n}\int_{\mathbb{R}^{3}}\sqrt{1 + a^{2}|\bar{p}|^{2}}R(n,
Q)u(\bar{p}) d\bar{p} = \int_{\mathbb{R}^{3}}\sqrt{1 +
a^{2}|\bar{p}|^{2}}u(\bar{p}) d\bar{p}
\end{equation}
If we make in each side of (\ref{eq:3.30}) the change of variables
$\bar{q} = B\bar{p}$ where \\
$B = Diag(a, a, a)$, then $|\bar{q}|^{2} = a^{2}|\bar{p}|^{2}$ ;
$d\bar{p} = \frac{1}{a^{3}}d\bar{q}$ and (\ref{eq:3.30}) gives,
using definition \ref{eq:2.4} of $\|.\|$:
\begin{equation}\label{eq:3.31}
\sqrt{n}\| R(n, Q) u o B^{-1}\| = \| u o B^{-1}\|
\end{equation}
But, if we compute $\| u o B\|$, using the above change of
variable, we have:
\begin{equation*}
\begin{aligned}
\| u o B\| &= \int_{\mathbb{R}^{3}}\sqrt{1 + |\bar{p}|^{2}}u o
B(\bar{p})d\bar{p} =  \int_{\mathbb{R}^{3}}\sqrt{1 +
\frac{1}{a^{2}}|\bar{q}|^{2}}|u(\bar{q})|\frac{1}{a^{3}}d\bar{q}\\
&\leq \frac{1}{a^{3}}\sqrt{1 +
\frac{1}{a^{2}}}\int_{\mathbb{R}^{3}}\sqrt{1 + |\bar{q}|^{2}}u
(\bar{q})d\bar{q}
\end{aligned}
\end{equation*}
i.e
\begin{equation*}
\| u o B\|  \leq \frac{1}{a^{3}} \left(1 + \frac{1}{a}\right)\| u
\|
\end{equation*}
The assumption $a \geq \frac{3}{2}$ implies
$\frac{1}{a^{3}}\left(1 + \frac{1}{a}\right)  \leq 1$, so that $\|
u o B\| \leq \| u \|$, this implies that $ u o B \in X_{r}$ if $u
\in X_{r}$. Now since (\ref{eq:3.31}) holds for every $u \in
X_{r}$, we have, replacing in (\ref{eq:3.31}) $u$ by $u o B$,
$\|\sqrt{n}R(n, Q)u\| = \| u \|$.
 We then have (\ref{eq:3.28}) and lemma 3.2 is proved.\\
 Now (\ref{eq:3.28}) is exactly equality (\ref{eq:3.10}) in
 proposition 3.3 in \cite{s}, we then prove exactly as for prop. 3.3 in \cite{s}, that all the other relations of that proposition hold in the present case. Using this result, the proof of point 2)a) of proposition 3.1 is the same as the proof of prop 4.1 in \cite{s}  and the proof of point 2) b) of
 prop 3.1 is the same as the proof of theorem 4.1 in \cite{s},
 just replacing, $[0, + \infty[$ by $[t_{0}, t_{0} + T]$. This
 completes the proof of prop 3.1 which give directly theorem 3.1.
 $\blacksquare$

 \section{Existence Theorem for the Einstein Equations}
 \subsection{Expression and Reduction of the Einstein Equations}
 We express the Einstein equations (\ref{eq:2.14}) in terms of the
 cosmological expansion factor a which is the only unknown. We
 have to compute the Ricci tensor $R_{\alpha\beta}$ given by
( \ref{eq:2.16}). \\
The expression (\ref{eq:2.12}) of $\Gamma_{\alpha\beta}^{\lambda}$
shows that the only non-zero components of the Ricci
 tensor  are the $R_{\alpha\alpha}$ and that $R_{11} = R_{22} =
 R_{33}$. Then, it will be enough to compute $R_{00} = R^{\lambda}_{\quad 0, \, \lambda 0
 }$and $R_{11} = R^{\lambda}_{\quad 1, \, \lambda 1}$. The expression
( \ref{eq:2.12}) of $\Gamma_{\alpha\beta}^{\lambda}$ and formulae
(\ref{eq:2.16}) give:
 \begin{equation*}
R^{0}_{0, \, 0 0} = 0; \quad R^{1}_{0, \, 1 0} = R^{2}_{0, \, 2 0}
= R^{3}_{0, \, 3 0} = - \frac{\ddot{a}}{a};
\end{equation*}
 \begin{equation*}
R^{0}_{1, \, 0 1} = a\ddot{a}; \quad R^{1}_{1, \, 1 1} = 0; \quad
R^{2}_{1, \, 2 1} = R^{3}_{1, \, 3 1} = (\dot{a})^{2}
\end{equation*}
We then deduce that:
 \begin{equation*}
R_{00} = - 3\frac{\ddot{a}}{a} \quad and \quad R_{11} = a\ddot{a}
+ 2(\dot{a})^{2}
\end{equation*}
We can then compute the scalar curvature R to be:
\begin{equation*}
R = R_{\alpha}^{\alpha} =  g^{\alpha\beta}R_{\alpha\beta} =
6\left[\frac{\ddot{a}}{a} +
\left(\frac{\dot{a}}{a}\right)^{2}\right]
\end{equation*}
The Einstein equations (\ref{eq:2.14}) then take the reduced form,
using expression (\ref{eq:2.1}) of g:
\begin{equation*}
\begin{cases}
\qquad 3(\frac{\dot{a}}{a})^{2} -  \Lambda  &= 8 \pi T_{00} \\
\\
- (\dot{a})^{2} - 2 a \ddot{a} + a^{2} \Lambda  &= 8 \pi T_{11}
\end{cases}
\end{equation*}
that can be written, using $T^{\alpha\beta} =
g^{\alpha\lambda}g^{\beta\mu}T_{\lambda\mu}$:
\begin{equation}\label{eq:4.1}
\qquad (\frac{\dot{a}}{a})^{2} = \frac{8 \pi}{3} T^{00}  +
\frac{\Lambda}{3}
\end{equation}
\begin{equation}\label{eq:4.2}
 \frac{\ddot{a}}{a} = - \frac{4 \pi}{3}(T^{00} +
3a^{2}T^{11}) + \frac{\Lambda}{3}
\end{equation}
in which $T^{\alpha\beta}$ is defined in term of f by
(\ref{eq:2.15}). In this paragraph, we suppose f fixed in $C[[0,
T]; X_{r}]$, with $f(0) = f_{0} \in L_{2}^{1}(\mathbb{R}^{3})$,
$f_{0} \geq 0$ a.e. and $r > \|f_{0}\|$.
\subsection{Compatibility}
The relations $R_{0i} = 0$, $R_{ij} = 0$ if $i \neq j$, $R_{11} =
R_{22} = R_{33}$, imply for the Einstein tensor $S_{\alpha\beta} =
R_{\alpha\beta} - \frac{1}{2}g_{\alpha\beta}R$, that:
\begin{equation}\label{eq:4.3}
T_{11} = T_{22} = T_{33}, \quad T_{0i} = 0, \quad  T_{ij} = 0
\quad for \quad i \neq j
\end{equation}
But the stress-matter tensor $T_{\alpha\beta}$ is defined by
(\ref{eq:2.15}) in terms of the distribution function f. So, the
relation (\ref{eq:4.3}) are in fact conditions to impose to f. we
prove:
\begin{proposition}
Let $f_{t_{0}}$ and $r > 0$ be defined as in theorem 3.1. Assume
that, in addition, $f_{t_{0}}$ is invariant by $S_{O_{3}}$ and
that, the collision kernel A satisfies
\begin{equation}\label{eq:4.4}
A(a(t), M\bar{p} ,  M\bar{q}, M\bar{p}' , M\bar{q}') = A(a(t) ,
\bar{p} , \bar{q} , \bar{p}' , \bar{q}') \,, \qquad \forall M \in
S_{O_{3}}
\end{equation}
then
\begin{itemize}
    \item[1)] The solution $f$ of the integral equation
    (\ref{eq:3.17})
    satisfies:
    \begin{equation}\label{eq:4.5}
f(t_{0} + t, M\bar{p}) = f(t_{0} + t, \bar{p})\,, \quad \forall t
\in [0, T], \quad \forall \bar{p} \in \mathbb{R}^{3}\,, \quad
\forall M \in S_{O_{3}}
\end{equation}
    \item[2)] The stress-matter tensor $T_{\alpha\beta}$
    satisfies the conditions (\ref{eq:4.3}).
\end{itemize}
\end{proposition}
\textbf{Proof}\\
1) Let $M \in S_{O_{3}}$; (\ref{eq:3.17}) gives, since
$f_{t_{0}}(M\bar{p}) = f_{t_{0}}(\bar{p})$ and $p^{0} =
p^{0}(\bar{p})$:
\begin{equation}\label{eq:4.6}
f(t_{0} + t , M\bar{p}) = f_{t_{0}}(\bar{p}) + \int_{t_{0}}^{t_{0}
+ t}\frac{1}{p^{0}oM}Q(f, f)(s, M\bar{p})ds , t \in [0, T]
\end{equation}
Notice that $p^{0} = p^{0}(\bar{p})$ is invariant by $S_{O_{3}}$.
Now, definition (\ref{eq:3.1}), (\ref{eq:3.2}), (\ref{eq:3.3}) of
Q gives:
\begin{equation}\label{eq:4.7}
Q(f, f)(s)(M\bar{p})=
\int_{\mathbb{R}^{3}}\frac{a^{3}(s)d\bar{q}}{q^{0}}\int_{S^{2}}[f(s,
\bar{p}')f(s, \bar{q}') - f(s, M\bar{p})f(s,\bar{q}) ]A(a ,
M\bar{p} , \bar{q} , \bar{p}' , \bar{q}')d\omega
\end{equation}
Let us set in (\ref{eq:4.7}) $\bar{q} = M\bar{q}_{1}$; $\omega =
M\omega_{1}$. Then formulae (\ref{eq:3.11}) give using expression
(\ref{eq:3.12}) of b, the invariance of the scalar product in
$\mathbb{R}^{3}$ by $S_{O_{3}}$:
\begin{equation*}
\left\{%
\begin{array}{ll}
 \bar{p}' = M\bar{p} + b(M\bar{p}, \bar{q}, \omega)\omega = M\bar{p} + b(M\bar{p}, M\bar{q}_{1}, M\omega_{1})M\omega_{1} = M\bar{p} + b(\bar{p}, \bar{q}_{1}, \omega_{1})M\omega_{1}  \\
 \bar{q}' = \bar{q} - b(M\bar{p}, \bar{q}, \omega)\omega = M\bar{q}_{1} - b(M\bar{p}, M\bar{q}_{1}, M\omega_{1})M\omega_{1} = M\bar{q}_{1} + b(\bar{p}, \bar{q}_{1}, \omega_{1})M\omega_{1}
\end{array}%
\right.
\end{equation*}
So that $\bar{p}' = M \bar{p}'_{1}$; $\bar{q}' = M \bar{q}'_{1}$
where:
\begin{equation*}
 \bar{p}'_{1} = \bar{p} + b(\bar{p}, \bar{q}_{1},
 \omega_{1})\omega_{1}; \quad  \bar{q}'_{1} = \bar{q}_{1} - b(\bar{p}, \bar{q}_{1},
 \omega_{1})\omega_{1}
\end{equation*}
Then , (\ref{eq:4.7}) implies, using assumption (\ref{eq:4.4}) on
A, $q^{0} = q_{1}^{0}$, and the invariance of the volume elements
$d\bar{q}$, $d\omega$ by $S_{O_{3}}$
\begin{equation*}
Q(f, f)(s)(M\bar{p}) = Q[f(s)oM, f(s)oM](\bar{p})
\end{equation*}
(\ref{eq:4.6}) then write:
\begin{equation*}
f(t_{0} + t)oM(\bar{p}) = f_{t_{0}}(\bar{p}) + \int_{t_{0}}^{t_{0}
+ t}\frac{1}{p^{0}}Q(f(s)oM, f(s)oM)(\bar{p})ds
\end{equation*}
which shows, by setting $h(s) = f(s)oM$, that $\|h(s)\| =
\|f(s)\|$  setting $\bar{q} = M\bar{p}$ in the integral in
$\bar{p}$ defining $\|f(s)oM\|$ , and that, h and f are 2
solutions in $C[[t_{0}, t_{0} + T]; X_{r}]$ of the integral
equation (\ref{eq:3.17}); the uniqueness theorem 3.1, then implies
h = f
and the first point in prop 4.1 is proved.\\
2) Let us consider the following element of $S_{O_{3}}$ in  which
$\theta \in \mathbb{R}^{3}$:
\begin{equation*}
 M_{1}(\theta) = \begin{pmatrix}
   _{cos\theta} & _{- sin\theta} & _{0} \\
   _{sin\theta} & _{cos\theta} & _{0} \\
   _{0} & _{0} & _{1} \
 \end{pmatrix}
\quad
 M_{2}(\theta) = \begin{pmatrix}
   _{cos\theta} & _{0} & _{- sin\theta} \\
   _{0} & _{1} & _{0} \\
   _{sin\theta} & _{0} & _{cos\theta} \
 \end{pmatrix}
\quad
 M_{3}(\theta) = \begin{pmatrix}
   _{1} & _{0} & _{0} \\
   _{0} & _{cos\theta} & _{-sin\theta} \\
   _{0} & _{sin\theta} & _{cos\theta} \
 \end{pmatrix}
 \end{equation*}
Consider the expression (\ref{eq:2.15}) of $T^{\alpha\beta}$ in
which f satisfies (\ref{eq:4.5}) and observe that
$p^{0}[M_{k}(\theta)\bar{q}] = q^{0}$, k = 1, 2, 3, $\theta \in
\mathbb{R}$.
\begin{itemize}
    \item[(i)] Set in (\ref{eq:2.15}) $\alpha = 0$, $\beta = i$ with
    i = 1, 2; now compute the integral using the change of
    variable $\bar{p} = M_{1}(\pi)\bar{q}$; the integral in
    $\bar{q}$ gives, using (\ref{eq:4.5}): $T^{0i} = - T^{0i}$, i =
    1, 2; hence $T^{01} = T^{02} = 0 $.
    \item[(ii)] Set in (\ref{eq:2.15}) $\alpha = 0$, $\beta = 3$ and compute the integral in $\bar{q}$ using the change of
    variable $\bar{p} = M_{3}(\pi)\bar{q}$; the integral in
    $\bar{q}$ gives, using (\ref{eq:4.5}): $T^{03} = - T^{03}$ hence $ T^{03} = 0 $.
    \item[(iii)] Set in (\ref{eq:2.15}) $\alpha = i$, $\beta = j$ $i \neq j$
    and compute the integral using the change of
    variable $\bar{p} = M_{k}(\frac{\pi}{2})\bar{q}$ with k = 1 if i = 1, j = 2; k = 2 if i = 1, j = 3, k = 3 if i = 2, j = 3 the integrals in
    $\bar{q}$ give, using (\ref{eq:4.5}): $T^{12} = - T^{12}$; $T^{13} = - T^{13}$; $T^{23} = - T^{23}$ hence $T^{12} = T^{13} = T^{23} = 0 $.
    \item[(iv)] Set in (\ref{eq:2.15}) $\alpha = \beta = i$ with
    i = 1, 2; and compute the integral using the change of
    variable $\bar{p} = M_{k}(\frac{\pi}{2})\bar{q}$; taking k = 1 if i = 1 and k = 3 if i = 2, to obtain using \ref{eq:4.5}: $T^{11} = T^{22}$ and  $T^{22} = T^{33}$.
\end{itemize}
This completes the proof of prop 4.1.

In all what follows, we assume that $f_{t_{0}}$ is invariant by
$S_{O_{3}}$ and that the collision kernel A satisfies assumption
(\ref{eq:4.4}). Notice that A defined in Remark 3.1 is an example
of such a kernel.
\subsection{The Constraint Equation}
We study the Cauchy problem for the system
(\ref{eq:4.1})-(\ref{eq:4.2}) on $[0, T]$ with initial data:
\begin{equation}\label{eq:4.8}
a(0) = a_{0}; \quad \dot{a}(0) = \dot{a}_{0}
\end{equation}
The Einstein equations (\ref{eq:2.14}) with $G = 1$, give, using
the Einstein tensor $S^{\alpha\beta}$:
\begin{equation*}
H^{0}_{\alpha} := S^{0}_{\alpha} + \Lambda g^{0}_{\alpha} - 8\pi
T^{0}_{\alpha} = 0
\end{equation*}
it is proved in \cite{u}, p. 39, that, in the general case, the
four quantities $H^{0}_{\alpha}$ do not involve the second
derivative of the metric tensor $g$ with respect to $t$, and,
using the identities $\nabla_{\alpha}(S^{\alpha\beta} - \Lambda
g^{\alpha\beta}  - 8 \pi T^{\alpha\beta}) = 0$, that the four
quantities $H^{0}_{\alpha}$ satisfy a linear homogeneous first
order differential system. consequently:
\begin{itemize}
    \item[1°)] For $t = 0$, the quantities $H^{0}_{\alpha}$
    express uniquely in term of the initial data $a_{0}$,
    $\dot{a}_{0}$and $f_{0}$.
    \item[2°)] If $H^{0}_{\alpha}(0) = 0$, then $H^{0}_{\alpha}(t) = 0$
    in the whole existence domain of the solution $g$ of the
    Einstein equations.
\end{itemize}
But $H^{0}_{\alpha}(0) = 0$ are 4 constraints to impose to the
initial data at $t = 0$ and the equation $H^{0}_{\alpha} = 0$ are
called \textbf{constraint equations}. In our case, the relations
$S^{0}_{i} = 0$, $T^{0}_{i} = 0$ (see prop 4.1) show that the
constraints $H^{0}_{i} = 0$, called the \textbf{momentum
constraints} are automatically satisfied. It then remains the
constraint with $\alpha = 0$, which is equivalent  to $S^{00} -
\Lambda - 8 \pi T^{00} = 0$, called the \textbf{Hamiltonian
constraint}, and which is
exactly (\ref{eq:4.1}).\\
So, following what we said above, the Hamiltonian constraint
(\ref{eq:4.1}) is satisfied in the whole existence domain of the
solution a of the initial values problem on $[0, T]$, if the
initial data $a_{0}$, $\dot{a}_{0}$, $f_{0}$ at $t = 0$, satisfy,
using expression (\ref{eq:2.15}) of $T^{00}$, the constraint
\begin{equation}\label{eq:4.10}
\left(\frac{\dot{a}_{0}}{a_{0}}\right)^{2} = \frac{8\pi
a_{0}^{3}}{3}\int_{\mathbb{R}^{3}}\sqrt{1 +
a_{0}^{2}|\bar{p}|^{2}}f_{0}(\bar{p})d\bar{p} + \frac{\Lambda}{3}
\end{equation}
(\ref{eq:4.10}) gives two possible choices of $\dot{a}_{0}$, when
$a_{0}$ and $f_{0}$ are given. We will choose, taking also into
account the hypothesis on $a(t)$ in theorem 3.1:
\begin{equation}\label{eq:4.9}
a_{0} \geq \frac{3}{2}; \quad f_{0} \in L^{1}_{2}(\mathbb{R}^{3});
\quad f_{0} \geq 0 \,\, a.e \quad \dot{a}_{0} > 0.
\end{equation}
We now concentrate on (\ref{eq:4.2}) which is the evolution
equation.
\subsection{The Evolution Equation}
We set $\theta = 3 \frac{\dot{a}}{a}$, then $\dot{\theta} = 3
[\frac{\ddot{a}}{a} - (\frac{\dot{a}}{a})^{2}]$ and \ref{eq:4.2}
gives:
\begin{equation}\label{eq:4.11}
\dot{\theta} = - \frac{\theta^{2}}{3} - 4 \pi(T^{00} +
3a^{2}T^{11}) + \Lambda
\end{equation}
(\ref{eq:4.11}) is the Raychaudhuri equation in $\theta$. We
prove:
\begin{proposition}
There can exist no global regular solution for the coupled
Einstein-Boltzmann system in the case $\Lambda < 0$
\end{proposition}
For the proof, we use the following result, proved in \cite{v}:
\begin{lemma}
Let u and $\theta$ be 2 differentiable functions of t satisfying:
\begin{equation}\label{eq:4.12}
      \dot{\theta} < - \frac{\theta^{2}}{3};  \quad
    \dot{u} = - \frac{u^{2}}{3}; \quad
    \theta(t_{1}) = u(t_{1})
  \end{equation}
\end{lemma}
for a given value $t_{1}$ of t. Then $\theta(t) \leq u(t)$ for $t
\geq t_{1}$. \\
\textbf{Proof of proposition 4.2}\\
Suppose $\Lambda < 0$. Then the Raychaudhuri equation
(\ref{eq:4.11}) gives $\dot{\theta} < - \frac{\theta^{2}}{3}$. We
have, integrating the equation in u on $[t_{1}, t]$ when $u \neq
0$ and since $\theta(t_{1}) = u(t_{1})$:
\begin{equation}\label{eq:4.13}
u(t) = \frac{3\theta(t_{1})}{3 + \theta(t_{1})(t - t_{1})}; \quad
t\geq t_{1}
\end{equation}
Let us show that we have necessary $\dot{a} < 0$.\\
The Hamiltonian constraint (\ref{eq:4.1}) writes, using
(\ref{eq:2.15}) and $p^{0} = \sqrt{1 + a^{2}(t)|\bar{p}|^{2}}$.
\begin{equation*}
\left(\frac{\dot{a}}{a}\right)^{2} = \frac{8\pi
a^{3}}{3}\int_{\mathbb{R}^{3}}\sqrt{1 + a^{2}|p|^{2}}f(t,
\bar{p})d\bar{p} + \frac{\Lambda}{3} \tag{a}
\end{equation*}
The derivative of the l.h.s is: $A =
2(\frac{\dot{a}}{a})\left[\frac{\ddot{a}}{a} -
\left(\frac{\dot{a}}{a}\right)^{2}\right]$. the derivative of the
r.h.s is
\begin{equation*}
B(t) =
\frac{8\pi}{3}\left[3a^{2}\dot{a}\int_{\mathbb{R}^{3}}p^{0}f(t,
\bar{p})d\bar{p} +
a^{3}\int_{\mathbb{R}^{3}}\frac{a\dot{a}|\bar{p}|^{2}}{p^{0}}f(t,
\bar{p})d\bar{p} + \int_{\mathbb{R}^{3}}p^{0}\frac{\partial
f}{\partial t}(t, \bar{p})d\bar{p}\right] \tag{b}
\end{equation*}
But we have by (\ref{eq:3.14}) $\frac{\partial f}{\partial t} =
\frac{Q(f, f)}{p^{0}}$, then (\ref{eq:3.29}) shows that the last
integral in B(t) is zero. We have, naturally, A(t) = B(t). Suppose
$\dot{a}(t_{1}) > 0$; (b) implies  $B(t_{1}) > 0$. Now
$\dot{\theta} < - \frac{\theta^{2}}{3} \Rightarrow \dot{\theta} <
0$. But $\theta = 3 \frac{\dot{a}}{a}$. Then $\dot{\theta} = 3
\left[\frac{\ddot{a}}{a} -
\left(\frac{\dot{a}}{a}\right)^{2}\right] < 0$, so that $A(t_{1})
< 0$ and the equality $A(t_{1}) = B(t_{1})$ is impossible. We then
have $\dot{a}(t_{1}) \leq 0$. The hypothesis $\dot{a}(t_{1}) = 0$
would implies that the r.h.s of (a) be a constant, but it is not
the case, since it depends on f which changes with $f_{t_{0}}$ We
then conclude that, necessarily $\dot{a} < 0$, which implies
$\theta < 0$, so that in (\ref{eq:4.12}) we have $\theta(t_{1}) <
0$. Now (\ref{eq:4.13}) shows that , since by (\ref{eq:4.12}) u is
a decreasing function:
\begin{equation}\label{eq:4.14}
u(t) \rightarrow -\infty \quad when \quad 3 + \theta(t_{1})(t -
t_{1}) \underset{>}{\rightarrow} 0, \quad and \quad t_{1} \leq t <
t_{1} - \frac{3}{\theta(t_{1})} := t^{*}
\end{equation}
By lemma 4.1, (\ref{eq:4.14}) implies that $\theta(t) = 3
\frac{\dot{a}}{a}(t) \rightarrow - \infty$ when $t
\underset{<}{\rightarrow} t^{*}$, then:
\begin{equation}\label{eq:4.15}
\left(\frac{\dot{a}(t)}{a(t)}\right)^{2}  \rightarrow + \infty
\quad when \quad t \underset{<}{\rightarrow} t^{*}
\end{equation}
Now, since $\dot{a} < 0$, a is a decreasing function on $[t_{1},
t^{*}[$, and then, $a(t) \leq a(t_{1})$, $\forall t \in [t_{1},
t^{*}[$; (a) then implies, since $\Lambda < 0$:
\begin{equation*}
(\dot{a})^{2} \leq \frac{8\pi}{3}a^{5}(t_{1}) \sqrt{1 +
a^{2}(t_{1})}\||f\|| := C^{2}(t_{1}, f)\,, \quad \forall t \in
[t_{1}, t^{*}[
\end{equation*}
so that:
\begin{equation*}
\left(\frac{\dot{a}(t)}{a(t)}\right)^{2}  \leq \frac{C^{2}(t_{1},
f)}{a^{2}(t)} \quad \forall t \in [t_{1}, t^{*}[
\end{equation*}
(\ref{eq:4.15}) then implies that $\frac{C^{2}(t_{1},
f)}{a^{2}(t)} \rightarrow +\infty$ when $t
\underset{<}{\rightarrow} t^{*}$ and this can happen only if:
$a(t) \rightarrow 0$ when $t \underset{<}{\rightarrow} t^{*}$. So
the cosmological expansion factor a tends to zero in a finite time
and such a solution (a, f) cannot be global towards the future.
This complete the proof of
proposition 4.2. $\blacksquare$\\
We now study the case $\Lambda > 0$. Notice that since $T^{00}
\geq 0$,  $T^{11} \geq 0$ (\ref{eq:4.1})-(\ref{eq:4.2}) gives by
subtraction $\frac{\ddot{a}}{a} - (\frac{\dot{a}}{a})^{2} < 0$;
this implies $\frac{d}{dt}(\frac{\dot{a}}{a}) < 0$, which shows
that, in all the cases, $\theta = 3 \frac{\dot{a}}{a}$ is a
decreasing function. In the case $\Lambda > 0$, the Hamiltonian
constraint (\ref{eq:4.1}) gives:
$\left(\frac{\dot{a}(t)}{a(t)}\right)^{2} \geq \frac{\Lambda}{3}$
i.e $\left[\frac{\dot{a}(t)}{a(t)} -
\sqrt{\frac{\Lambda}{3}}\right]\left[\frac{\dot{a}(t)}{a(t)} +
\sqrt{\frac{\Lambda}{3}})\right] \geq 0$, which is equivalent to:
\begin{equation}\label{eq:4.16}
\frac{\dot{a}(t)}{a(t)} \geq \sqrt{\frac{\Lambda}{3}}
\end{equation}
or
\begin{equation}\label{eq:4.17}
\frac{\dot{a}(t)}{a(t)} \leq - \sqrt{\frac{\Lambda}{3}}
\end{equation}
The continuity of $t \, \rightarrow \, \frac{\dot{a}(t)}{a(t)}$
implies that we have to choose between (\ref{eq:4.16}) and
(\ref{eq:4.17}). Since $a > 0$, (\ref{eq:4.17}) implies $\dot{a} <
0$ and a is decreasing; (\ref{eq:4.16}) implies that $\dot{a} >
0$, then a is increasing and since $\frac{\dot{a}}{a}$ is
decreasing, this gives on $[t_{0}, t]$:
\begin{equation}\label{eq:4.18}
a(t) \geq a(t_{0}); \quad \sqrt{\frac{\Lambda}{3}}\, \leq \,
\frac{\dot{a}(t)}{a(t)}\, \leq \, \frac{\dot{a}(t_{0})}{a(t_{0})},
\quad t \geq t_{0}
\end{equation}
Recall that our aim is to study the coupled Einstein-Boltzmann
system and we had to require, for the study of the Boltzmann
equation, that a, which is positive be bounded away from zero.
This problem is solved by choosing (\ref{eq:4.16}). Another
important consequence of (\ref{eq:4.16}) is that it implies on
$[t_{0}, t]$
\begin{equation*}
a(t) \geq a(t_{0})e^{\sqrt{\frac{\Lambda}{3}}(t - t_{0})}, \quad t
\geq t_{0}
\end{equation*}
Which shows that, the cosmological expansion factor has an
exponential growth; but, it also shows that, an eventual global
solution a will be unbounded, and this is why, in order to use
standard results, we make the change of variable:
\begin{equation}\label{eq:4.19}
e = \frac{1}{a}
\end{equation}
which gives:
\begin{equation}\label{eq:4.20}
\dot{e} = - \frac{\dot{a}}{a^{2}}
\end{equation}
Then, using \ref{eq:2.15}, $T^{00}$ and $T^{11}$ in the r.h.s of
\ref{eq:4.2} express in, terms of e and f, and we set:
\begin{equation}\label{eq:4.21}
\rho = T^{00}  = \frac{1}{e^{3}}\int_{\mathbb{R}^{3}}\sqrt{1 +
\frac{1}{e^{2}}|\bar{p}|^{2}}f(t, \bar{p})d\bar{p}
\end{equation}
\begin{equation}\label{eq:4.22}
P = a^{2}T^{11}  =
\frac{1}{e^{5}}\int_{\mathbb{R}^{3}}\frac{(p^{1})^{2}f(t,
\bar{p})}{\sqrt{1 + \frac{1}{e^{2}}|\bar{p}|^{2}}}d\bar{p}
\end{equation}
$\rho$ stands for the density and P for the pressure. One verifies
that:
\begin{equation}\label{eq:4.23}
P \leq \rho
\end{equation}
Recall that $r > 0$ is such that $r > \|f_{0}\|$. If we set:
\begin{equation}\label{eq:4.24}
d_{0} = 3 \sqrt{\frac{\Lambda}{3} +  \frac{16\pi}{3}r a_{0}^{4}}
\end{equation}
One checks easily, using $a \geq a_{0} \geq \frac{3}{2}$,
(\ref{eq:4.18}) with $t_{0} = 0$ (\ref{eq:4.9}) (\ref{eq:4.10})
and (\ref{eq:4.24}), that
\begin{equation}\label{eq:4.25}
e = \frac{1}{a} \in ]0, \frac{2}{3}] ;\quad et \quad \theta =
3\frac{\dot{a}}{a} \in [\sqrt{3\Lambda} \,,\,d_{0}]
\end{equation}
A direct calculation shows, using (\ref{eq:4.19}),
(\ref{eq:4.20}), (\ref{eq:4.21}), (\ref{eq:4.22}) , and
(\ref{eq:4.11}), that the Einstein evolution equation
(\ref{eq:4.2}) is equivalent to the  following first order system
in $(e, \theta)$:\\
\begin{equation}\label{eq:4.26}
\dot{e} = - \frac{\theta}{3}\times e
\end{equation}
\begin{equation}\label{eq:4.27}
\dot{\theta} = - \frac{\theta^{2}}{3} - 4\pi(\rho + 3 P) + \Lambda
\end{equation}
with $\rho = \rho(e, f)$ and $P = P(e, f)$ given by
(\ref{eq:4.21}) and (\ref{eq:4.22}).\\
with will study study the initial values problem for the system
(\ref{eq:4.26})-(\ref{eq:4.27}) with initial data $(e_{0},
\theta_{0})$ and $t = 0$. By virtue of the change of variables
(\ref{eq:4.25}), we will deduce solution for the Einstein
evolution equation (\ref{eq:4.2}) by setting:
\begin{equation}\label{eq:4.28}
e(0) = \frac{1}{a_{0}} \, \quad \theta(0) =
3\frac{\dot{a}_{0}}{a_{0}}
\end{equation}
in which $a_{0}$, $\dot{a}_{0}$ satisfy the constraint
(\ref{eq:4.9})-(\ref{eq:4.10})with $f_{0}$ given. By virtue of
(\ref{eq:4.25}) it will be enough for the study of the initial
value problem for (\ref{eq:4.26})-(\ref{eq:4.27}) to take $e_{0}$,
$\theta_{0}$ such that
\begin{equation}\label{eq:4.28'}
0 < e_{0} \leq \frac{2}{3}, \quad 0 < \theta_{0} \leq d_{0}
\end{equation}
\begin{remark}
The equivalence of the evolution equation (\ref{eq:4.2}) and the
system (\ref{eq:4.26})-(\ref{eq:4.27}) requires that any solution
$(e, \theta)$ of (\ref{eq:4.26})-(\ref{eq:4.27})-(\ref{eq:4.28'})
satisfies $e > 0$ and $\theta > 0$.
\end{remark}
In fact, (\ref{eq:4.21}), (\ref{eq:4.22}) require $e \neq 0$; by
(\ref{eq:4.28'}), $e(0) = e_{0} > 0$. Since $e$ is continuous and
cannot vanish, by virtue of the mean values problem, $e$ remains
strictly positive. Now if we have $\theta(t_{1}) = 0$, then
(\ref{eq:4.26}) implies $\dot{e}(t_{1}) = 0$ and (\ref{eq:4.20})
implies $\dot{a}(t_{1}) = 0$, which is impossible given
(\ref{eq:4.16}). So $\theta$ never vanishes and by
(\ref{eq:4.28'}) $\theta(0) = \theta_{0} > 0$; So $\theta > 0$.\\
For the global existence theorem, we will need the following a
priori estimation
\begin{proposition}
Let $\delta > 0$ and $t_{0} \in \mathbb{R}_{+}$ be given; suppose
that in (\ref{eq:4.21}) and (\ref{eq:4.22}), $f \in C[t_{0}, t_{0}
+ \delta; L^{1}_{2}(\mathbb{R}^{3})]$ is given. Let $(e, \theta)$
be any solution of the system (\ref{eq:4.26})-(\ref{eq:4.27}) on
$[t_{0}, t_{0} + \delta]$. Then $\theta$ and $a = \frac{1}{e}$
satisfy the inequalities:
\begin{equation}\label{eq:4.281}
\theta(t_{0} + t) \leq \theta(t_{0})+ \Lambda, \quad t \in [0,
\delta]
\end{equation}
\begin{equation}\label{eq:4.282}
a(t_{0} + t) \leq a(t_{0})e^{(\frac{\theta(t_{0})}{3} +
\frac{\Lambda}{3})(t_{0} + t + 1)^{2}} , \quad t \in [0, \delta]
\end{equation}
\end{proposition}
\textbf{Proof}.\\
Since $\rho \geq 0$, $P \geq 0$, (\ref{eq:4.27}) implies:
$\dot{\theta} \leq \Lambda$. Integrating this inequality on
$[t_{0}, t_{0} + t]$ where $t \in [0, \delta]$ yields
(\ref{eq:4.281}).\\
Next, since $e > 0$, integrating (\ref{eq:4.26}) that writes $-
\frac{\dot{e}}{e} = \frac{\theta}{3}$ over $[t_{0}, t_{0} + t]$,
$t \in [0, \delta]$ gives:
\begin{equation}\label{eq:a}
a(t_{0} + t) \leq a(t_{0})e^{\int_{t_{0}}^{t_{0} +
t}\frac{\theta(s)}{3}ds} , \quad t \in [0, \delta]
\end{equation}
Now setting in (\ref{eq:4.281}) $s = t_{0} + t \in [t_{0}, t_{0} +
\delta]$,and integrating on $[t_{0}, t_{0} + t]$ yields:
\begin{equation}\label{eq:b}
\int_{t_{0}}^{t_{0} + t}\frac{\theta(s)}{3}ds \leq
(\frac{\theta(t_{0})}{3} + \frac{\Lambda}{3})(t_{0} + t + 1)^{2} ,
\quad t \in [0, \delta]
\end{equation}
(\ref{eq:4.282}) then follows from (a) and (b).$\blacksquare$
 We deduce:
\begin{proposition}
Let $T > 0$ and $f \in [[0, T]; X_{r}]$ be given. Suppose that the
initial value problem
(\ref{eq:4.26})-(\ref{eq:4.27})-(\ref{eq:4.28}) with initial data
(\ref{eq:4.28}) at $t = 0$ satisfying the constraints
(\ref{eq:4.9})-(\ref{eq:4.10}) has a solution $(\Xi =
\frac{1}{\Omega}, \Theta)$ on $[0, t_{0}]$ with $0 \leq t_{0} <
T$. Then, any solution $(e = \frac{1}{a}, \theta)$ of the initial
value problem for the system (\ref{eq:4.26})-(\ref{eq:4.27}) on
$[t_{0}, t_{0} + \delta]$, $\delta > 0$, with initial data $(e,
\theta)(t_{0}) = (\Xi, \Theta)(t_{0})$ at $t = t_{0}$, satisfy the
inequalities:
\begin{equation}\label{eq:4.283}
a(t_{0} + t) \leq C(2)e^{C_{3}(t_{0} + t + 1)^{2}} , \quad t \in
[0, \delta]
\end{equation}
\begin{equation}\label{eq:4.284}
\theta(t_{0} + t) \leq 3\gamma_{1} + \Lambda(T + t) , \quad t \in
[0, \delta]
\end{equation}
where:
\begin{equation}\label{eq:4.285}
C_{2} = a_{0}e^{\gamma_{1}}; \quad C_{3} = \gamma_{1} +
\frac{\Lambda}{3}T; \quad \gamma_{1} = \gamma_{1}(a_{0}, r, T) =
(\frac{\Lambda}{3} + \sqrt{\frac{\Lambda}{3} + 3ra_{0}^{4}})(T +
1)^{2}
\end{equation}
\end{proposition}
\textbf{Proof}\\
We apply proposition 4.3 to the solution $(e, \theta)$ of
(\ref{eq:4.26})-(\ref{eq:4.27}) on $[t_{0}, t_{0} + \delta]$;
(\ref{eq:4.282}) gives, since $a(t_{0}) = \Xi(t_{0})$,
$\theta(t_{0}) = \Theta(t_{0})$:
\begin{equation*}
a(t_{0} + t) \leq \Xi(t_{0})e^{(\frac{\Theta(t_{0})}{3} +
\frac{\Lambda}{3})(t_{0} + t + 1)^{2}} , \quad t \in [0, \delta]
\tag{a}
\end{equation*}
Now apply (\ref{eq:4.282}) to the solution $(\Xi, \Theta)$ of
(\ref{eq:4.26})-(\ref{eq:4.27}) on $[0, t_{0}]$, at the point
$t_{0}$. We obtain by setting (\ref{eq:4.282}), $t_{0} = 0$, $t =
t_{0}$ and since $a(t_{0}) = \Xi(t_{0})$, $\Xi(0) = a_{0}$,
$\Theta(0) = 3\frac{\dot{a}_{0}}{a_{0}}$, $0 \leq t_{0} < T$:
\begin{equation*}
\Xi(t_{0}) \leq a_{0}e^{(\frac{\dot{a}_{0}}{a_{0}} +
\frac{\Lambda}{3})(t_{0} + 1)^{2}} \leq
a_{0}e^{(\frac{\dot{a}_{0}}{a_{0}} + \frac{\Lambda}{3})(T +
1)^{2}}, \quad t \in [0, \delta] \tag{b}
\end{equation*}
Now since the initial data at $t = 0$ satisfy (\ref{eq:4.9}),
(\ref{eq:4.10}), this imply, using $\dot{a}_{0} > 0$, $a_{0} > 1$,
$\|f_{0}\| < r$:
\begin{equation*}
\frac{\dot{a}_{0}}{a_{0}} \leq \sqrt{\frac{\Lambda}{3} + \frac{8
\pi}{3} r a_{0}^{4}} \tag{c}
\end{equation*}
and (b) gives:
\begin{equation*}
\Xi(t_{0}) \leq a_{0}e^{\gamma_{1}} \tag{d}
\end{equation*}
with
\begin{equation*}
\gamma_{1} = \left(\frac{\Lambda}{3} + \sqrt{\frac{\Lambda}{3} +
\frac{8 \pi}{3} r a_{0}^{4}}\right)(T + 1)^{2} \tag{e}
\end{equation*}
Now, we apply (\ref{eq:4.281}) to the solution $(\Xi, \Theta)$ of
(\ref{eq:4.26})-(\ref{eq:4.27})-(\ref{eq:4.28}), on $[0, t_{0}]$
at point $t_{0}$, $t = t_{0}$ and since $\Xi(0) = 3
\frac{\dot{a}_{0}}{a_{0}}$, $0 \leq t_{0} < T$:
\begin{equation*}
\Xi(t_{0}) \leq \Xi(0) + \Lambda t_{0} \leq 3
\frac{\dot{a}_{0}}{a_{0}} + \Lambda T \tag{f}
\end{equation*}
We obtain, using (c) and (f):
\begin{equation*}
\frac{\Xi(t_{0})}{3} + \frac{\Lambda}{3} \leq
\left(\frac{\Lambda}{3} + \sqrt{\frac{\Lambda}{3} + \frac{8
\pi}{3} r a_{0}^{4}}\right) + \frac{\Lambda}{3}T
\end{equation*}
This gives, using the definition (e) of $\gamma_{1}$:
\begin{equation*}
\frac{\Xi(t_{0})}{3} + \frac{\Lambda}{3} \leq \gamma_{1} +
\frac{\Lambda}{3}T \tag{g}
\end{equation*}
and (\ref{eq:4.283}) follows from (a), (d) and (g).\\
The inequality (\ref{eq:4.284}) follows from (\ref{eq:4.281}),
$\theta(t_{0}) = \Xi(t_{0})$, (f), (c) and (e). This completes the
proof of prop. 4.4. $\blacksquare$ \\
We can now prove.
\begin{proposition}
Let $T > 0$ and $f \in C[[0, T]; X_{r}]$ be given. Then the
initial value problem for the system
(\ref{eq:4.26})-(\ref{eq:4.27}) with initial data $(e_{0},
\theta_{0})$ at $t = 0$ satisfying (\ref{eq:4.28'}), has an unique
solution $(e, \theta)$ on $[0, T]$
\end{proposition}
The proof of proposition 4.5 will use the following result.
\begin{lemma}
Let $e_{1} ,  \,e_{2}\,  \in \, ]0 , \frac{2}{3}]$; then we have,
C being a constant:
\begin{equation}\label{eq:4.29}
|\rho(e_{1} , f) - \rho(e_{2} , f)| \leq
\frac{C}{e_{1}^{4}e_{2}^{6}}\|f(t)\||e_{1} - e_{2} |
\end{equation}
\begin{equation}\label{eq:4.30}
|P(e_{1} , f) - P(e_{2} , f)| \leq
\frac{C}{e_{1}^{4}e_{2}^{7}}\|f(t)\||e_{1} - e_{2} |
\end{equation}
\end{lemma}
\textbf{Proof of the lemma 4.2}\\
1) To prove, (\ref{eq:4.29}), we write (\ref{eq:4.21}) in $e_{1} ,
\, e_{2}\,  \in \, ]0 , \frac{2}{3}]$ and we subtract. We will
use:
\begin{equation*}
\left|\frac{1}{e_{1}^{3}} - \frac{1}{e_{2}^{3}} \right| =
\frac{|e_{1}^{3} - e_{2}^{3}|}{e_{1}^{3} e_{2}^{3}} =
  \frac{|e_{1} - e_{2}|}{e_{1}^{3} e_{2}^{3}}|e_{2}^{2} +
e_{1}e_{2} +e_{1}^{2} | \leq 3 \frac{|e_{1} - e_{2}|}{e_{1}^{3}
e_{2}^{3}} \tag{a}
\end{equation*}
We will also use, applying $|\sqrt{a} - \sqrt{b}| = |a -
b||\sqrt{a} + \sqrt{b}|^{-1}$, $0 < e_{i} < 1$ and (a):
\begin{equation*}
\begin{aligned}
 \left|\sqrt{1 + \frac{1}{e_{1}^{2}}|\bar{p}|^{2}} -
\sqrt{1 + \frac{1}{e_{2}^{2}}|\bar{p}|^{2}}\right| &\leq
\left|\frac{1}{e_{1}^{3}} -
\frac{1}{e_{2}^{3}}\right|\frac{|\bar{p}|^{2}}{\sqrt{1 +
\frac{|\bar{p}|^{2}}{e_{1}^{2}}}} \\
&\leq \frac{3\sqrt{1 + |\bar{p}|^{2}}|e_{1} -
e_{2}|}{e_{1}^{3}e_{2}^{3}}
\end{aligned}\tag{b}
\end{equation*}
Now (\ref{eq:4.21}) gives:
\begin{equation*}
\begin{aligned}
|\rho(e_{1} , \bar{f}) - \rho(e_{2} , \bar{f})| &\leq
\left|\frac{1}{e_{1}^{3}} -
\frac{1}{e_{1}^{3}}\right|\int_{\mathbb{R}^{3}}\sqrt{1 +
\frac{1}{e_{1}^{2}}|\bar{p}|^{2}}f(t, \bar{p})d\bar{p} \\
&+ \frac{1}{e_{2}^{3}}\int_{\mathbb{R}^{3}}\left|\sqrt{1 +
\frac{1}{e_{1}^{2}}|\bar{p}|^{2}} - \sqrt{1 +
\frac{1}{e_{2}^{2}}|\bar{p}|^{2}}\right|f(t, \bar{p})d\bar{p}
\end{aligned}\tag{c}
\end{equation*}
and (\ref{eq:4.29}) follows from (a), (b) (c) and $e_{i} \leq 1$.\\

2) To prove, (\ref{eq:4.30}), we write (\ref{eq:4.22}) in $e_{1} ,
\,e_{2}\,  \in \, ]0 , \frac{2}{3}]$ and we subtract. We will use:
\begin{equation*}
\left|\frac{1}{e_{1}^{5}} - \frac{1}{e_{2}^{5}} \right| =
\frac{|e_{1}^{5} - e_{2}^{5}|}{e_{1}^{5} e_{2}^{5}} =
 =  \frac{|e_{1} - e_{2}|}{e_{1}^{5} e_{2}^{5}}|e_{1}^{4} +
e_{1}^{3}e_{2} + e_{1}^{2}e_{2}^{2} + e_{1}e_{2}^{3} + e_{2}^{4}|
\leq 5 \frac{|e_{1} - e_{2}|}{e_{1}^{5} e_{2}^{5}} \tag{d}
\end{equation*}
and we also use $0 < e_{i} < 1$ and
$\frac{(p^{1})^{2}}{|\bar{p}|^{2}} \leq 1$ to obtain:
\begin{equation*}
\begin{aligned}
\left|\frac{(p^{1})^{2}}{\sqrt{1 +
\frac{1}{e_{1}^{2}}|\bar{p}|^{2}}} - \frac{(p^{1})^{2}}{\sqrt{1 +
\frac{1}{e_{2}^{2}}|\bar{p}|^{2}}}\right| &\leq \left|\sqrt{1 +
\frac{1}{e_{1}^{2}}|\bar{p}|^{2}} - \sqrt{1 +
\frac{1}{e_{2}^{2}}|\bar{p}|^{2}}\right|\frac{(p^{1})^{2}}{|\bar{p}|^{2}}\\
&\leq \left|\frac{1}{e_{1}^{2}} - \frac{1}{e_{1}^{2}}
\right|\frac{(\bar{p})^{2}}{\sqrt{1 +
\frac{1}{e_{2}^{2}}|\bar{p}|^{2}}}\\
&\leq \frac{2}{e_{2}^{2}}\sqrt{1 +
\frac{1}{e_{1}^{2}}|\bar{p}|^{2}}|e_{1} - e_{2}| \leq
\frac{2}{e_{1}e_{2}^{2}}\sqrt{1 + |\bar{p}|^{2}}|e_{1} - e_{2}|
\end{aligned} \tag{e}
\end{equation*}
Now (\ref{eq:4.21}) gives:
\begin{equation*}
\begin{aligned}
|P(e_{1} , f) - P(e_{2} , f)| &\leq \left|\frac{1}{e_{1}^{5}} -
\frac{1}{e_{2}^{5}}\right|\int_{\mathbb{R}^{3}}\frac{(p^{1})^{2}}{\sqrt{1
+ \frac{1}{e_{1}^{2}}|\bar{p}|^{2}}}f(t, \bar{p})d\bar{p} \\
&+
\frac{1}{e_{2}^{5}}\int_{\mathbb{R}^{3}}\left|\frac{(p^{1})^{2}}{\sqrt{1
+ \frac{1}{e_{1}^{2}}|\bar{p}|^{2}}} - \frac{(p^{2})^{2}}{\sqrt{1
+ \frac{1}{e_{2}^{2}}|\bar{p}|^{2}}}\right|f(t, \bar{p})d\bar{p}
\end{aligned} \tag{f}
\end{equation*}
(\ref{eq:4.30}) then follows from (d),(e), (f), $0 < e_{i} < 1$,
and using in the first integral:
\begin{equation}\label{eq:4.31}
\frac{(p^{1})^{2}}{\sqrt{1 + \frac{1}{e_{1}^{2}}|\bar{p}|^{2}}}
\leq e_{1}^{2}\sqrt{1 + \frac{1}{e_{1}^{2}}|\bar{p}|^{2}} \leq
e_{1}\sqrt{1 + |\bar{p}|^{2}}
\end{equation}
This completes the proof of the lemma 4.2. $\blacksquare$\\
\textbf{Proof of proposition 4.5}\\
Define using (\ref{eq:4.26}), (\ref{eq:4.27}), (\ref{eq:4.28'}),
the function $F$ on $]0 , \frac{2}{3}]\times]0, d_{0} ]$ by:
\begin{equation}\label{eq:4.32}
F(e , \theta) = \left[- \frac{\theta e}{3} ,\quad -
\frac{\theta^{2}}{3} - 4\pi (\rho + 3P) + \Lambda\right]
\end{equation}
To prove the proposition 4.5, we first prove the existence of a
local solution. To have that result, we have to show that F
defined by (\ref{eq:4.32}) is continuous with respect to t, and
locally Lipschitzian in $(e,
\theta)$ with respect to the norm of $\mathbb{R}^{2}$.\\
The dependence of F on t is through f that appears in $\rho$ and P
(see formulae (\ref{eq:4.21}) and (\ref{eq:4.22})). But, since $f
\in C[t_{0}, t_{0} + T, X_{r}]$, the definition (\ref{eq:2.6}) of
this space shows that F is a continuous function of t. Now we
have, with \\
$(e_{1}, \theta_{1}) \, ,\, (e_{2}, \theta_{2}) \, \in \, ]0 ,
\frac{2}{3}]\times]0, d_{0} ]$, and by usual factorization:
\begin{equation}\label{eq:4.33}
\left|\frac{\theta_{1}e_{1}}{3} - \frac{\theta_{2}e_{2}}{3}
\right| \leq \frac{1 + d_{0}}{3}(|\theta_{1} - \theta_{2}| +
|e_{1} - e_{2}|)
\end{equation}
\begin{equation}\label{eq:4.34}
\left|\frac{1}{3}(\theta_{1}^{2} - \theta_{2}^{2})\right| \leq
\frac{2d_{0}}{3}|\theta_{1} - \theta_{2}|
\end{equation}
Let us fix $e^{1} \, \in \, ]0 , \frac{2}{3}]$ and we take $e_{1}
, e_{2} \in \left]\frac{e^{1}}{2} , \frac{e^{1} +
\frac{2}{3}}{2}\right[$ then : $\frac{1}{e_{i}} \, \leq \,
\frac{2}{e^{1}}$ $i = 1, 2$ and we deduce from
(\ref{eq:4.29})-(\ref{eq:4.30}) that
\begin{equation}\label{eq:4.35}
|\rho(e_{1} , f) - \rho(e_{2} , f)| + 3|P(e_{1} , f) - P(e_{2} ,
f)| \leq N(e^{1})r|e_{1} - e_{2} |
\end{equation}
Where N is a constant depending only on $e^{1}$ .\\
Then, if $\mathbb{R}^{2}$ is endowed with the norm $\|(x,
y)\|_{\mathbb{R}^{2}} = |x| + |y|$, we have for F defined by
(\ref{eq:4.32}), using
(\ref{eq:4.33})-(\ref{eq:4.34})-(\ref{eq:4.35}):
\begin{equation}\label{eq:4.36}
 \| F(e_{1} , \theta_{1})  - F(e_{2} , \theta_{2})\|_{\mathbb{R}^{2}} \leq M(|\theta_{1} -
\theta_{2}| + |e_{1} - e_{2}|) = M\|(e_{1} , \theta_{1})  - (e_{1}
, \theta_{1})\|_{\mathbb{R}^{2}}
\end{equation}
Where M is a constant depending only on $e^{1}$, and $d_{0}$.\\
(\ref{eq:4.36}) shows that F is locally Lipschitzian in $(e,
\theta)$ with respect to the norm of $\mathbb{R}^{2}$. The
standard theorem of the first order differential system implies
that the initial value problem
(\ref{eq:4.26})-(\ref{eq:4.27})-(\ref{eq:4.28'}) has a unique
local solution $(e, \theta)$ on $[0, \delta]$, $\delta > 0$.
Notice that since $f \in C[[0, T]; X_{r}]$, the system
(\ref{eq:4.26})-(\ref{eq:4.27}) is defined on $[0, T]$. Hence the
maximum value of $\delta$ is $\delta = T$. Now if $0 < \delta <
T$, prop. 4.5 in which we set $t_{0} = 0$, $0 \leq t \leq \delta <
T$, shows, applying (\ref{eq:4.282}) and $a = \frac{1}{e}$, that
for every solution $e$, the function $\frac{1}{e}$ is uniformly
bounded, and, $\rho$ and $P$ defined by (\ref{eq:4.21}) and
(\ref{eq:4.22}) are uniformly bounded. Consequently, the function
$F$ defined by (\ref{eq:4.32}) is uniformly bounded. The standard
theorem on the first order system then implies that the solution
$(e, \theta)$ is global on $[0, T]$. This complete the  proof of
prop. 4.5. $\blacksquare$\\
We deduce a result that will be useful to prove the global
existence for the coupled Einstein-Boltzmann system. We will use
the number $D_{0}$ defined by:
\begin{equation}\label{eq:4.37}
D_{0} = 3 \gamma_{1} + \Lambda (T + 1)
\end{equation}
where $\gamma_{1}$ is defined by (\ref{eq:4.285}). We prove:
\begin{proposition}
Let $T > 0$ and $f \in C[[0, T]; X_{r}]$ be given. Let $(\Xi =
\frac{1}{\Omega}, \Theta)$ be the solution of the initial value
problem for the system (\ref{eq:4.26})-(\ref{eq:4.27}) with
initial data $(e_{0}, \theta_{0})$ at $t = 0$ given by
(\ref{eq:4.28}) with $a_{0}$, $\dot{a}_{0}$ satisfying the
constraints (\ref{eq:4.9})-(\ref{eq:4.10}). Let $t_{0} \in [0,
T]$. Then the initial values problem for the system
(\ref{eq:4.26})-(\ref{eq:4.27}), with the initial data $(\Xi =
\frac{1}{\Omega}, \Theta)(t_{0})$ at $t = t_{0}$ has a unique
solution $(e = \frac{1}{a}, \theta)$ on $[t_{0}, t_{0} + \delta]$,
where $\delta > 0$ is independent  of $t_{0}$. The solution $(e =
\frac{1}{a}, \theta)$ satisfies the inequalities:
\begin{equation}\label{eq:4.38}
\frac{3}{2} \leq a(t_{0} + t) \leq C_{2}e^{C_{3}(t_{0} + t +
1)^{2}}, \quad t \in [0, \delta]
\end{equation}
\begin{equation}\label{eq:4.39}
\sqrt{3\Lambda} \leq \theta(t_{0} + t) \leq 3 \gamma_{1} + \Lambda
(T + t), \quad t \in [0, \delta]
\end{equation}
where $C_{2}$, $C_{3}$ and $\gamma_{1}$ are defined by
(\ref{eq:4.285})
\end{proposition}
\textbf{Proof}\\
We have $\theta_{0} = 3 \frac{\dot{a}_{0}}{a_{0}}$ and
(\ref{eq:4.9}) implies $\theta_{0} \geq \sqrt{3\Lambda}$, so, the
proof given for prop. 4.5 with the function $F$ given by
(\ref{eq:4.32}) and defined this time on $]0, \frac{2}{3}] \times
[\sqrt{3\Lambda}, d_{0}]$ leads to a solution $(\Xi, \Theta)$
satisfying $\Theta \geq \sqrt{3\Lambda}$, hence, $\Theta(t_{0})
\geq \sqrt{3\Lambda}$. Now suppose that we look for solutions $(e,
\theta)$ on $[t_{0}, t_{0} + \delta]$ with $0 < \delta < 1$; then
(\ref{eq:4.284}) shows that every solution $(e, \theta)$ satisfies
$\theta(t_{0} + t) \leq D_{0}$, $0 < t < \delta$, where $D_{0}$ is
defined by (\ref{eq:4.37}). We prove the existence of a local
solution $(e, \theta)$ of (\ref{eq:4.26})-(\ref{eq:4.27}) on
$[t_{0}, t_{0} + \delta]$, with initial data at $t = t_{0}$:
$e(t_{0}) = \Xi(t_{0})$, $\theta(t_{0}) = \Theta(t_{0}) \geq
\sqrt{3\Lambda}$, following the same lines as in the proof of
prop. 4.5, with the function $F$ given by (\ref{eq:4.32}), defined
this time on $]0, \frac{2}{3}]\times[\sqrt{3\Lambda}, D_{0}]$.
This leads to the existence of a local solution $(e, \theta)$ on
some interval $[t_{0}, t_{0} + \delta]$, $0 < \delta < 1$. On the
other hand, (\ref{eq:4.283}) gives, using $0 \leq t_{0} < T$, $0 <
\delta < 1$: $a(t_{0} + t) = \frac{1}{e(t_{0} + t)} \leq
C_{2}e^{C_{3}(T + 2)^{2}}$, which shows that, for every solution
$(e, \theta)$, $\frac{1}{e}$ is uniformly bounded. From there, the
existence of a number $0 < \delta < 1$, independent of $t_{0}$.
Finally, (\ref{eq:4.38}) follows from (\ref{eq:4.283}) and
$\frac{1}{a(t_{0} + t)} = e(t_{0} + t) \leq \frac{3}{2}$;
(\ref{eq:4.39}) follows from (\ref{eq:4.284}) and what we said
above. This ends the proof of prop. 4.6. $\blacksquare$
\begin{theorem}
Let $T > 0$ and $f \in C[[0, T]; X_{r}]$ be given. Then the
initial value problem for the Einstein equation
(\ref{eq:4.1})-(\ref{eq:4.2})-(\ref{eq:4.8}) with initial data
$a_{0}$, $\dot{a}_{0}$ satisfying the constraints
(\ref{eq:4.9})-(\ref{eq:4.10}) has an unique solution $a$ on $[0,
T]$.
\end{theorem}
\textbf{Proof}\\
Apply proposition 4.5, choosing the initial data $(e_{0},
\theta_{0})$ at $t = 0$ by (\ref{eq:4.28}); (\ref{eq:4.25}) shows
that $(e_{0}, \theta_{0})$ satisfy (\ref{eq:4.28'}). Prop. 4.5
then proves the existence of the unique solution $(e, \theta)$ of
(\ref{eq:4.26})-(\ref{eq:4.27})-(\ref{eq:4.28}) on $[0, T]$. Now
the equivalence of the evolution equation (\ref{eq:4.2}) with the
system (\ref{eq:4.26})-(\ref{eq:4.27}) show that $a = \frac{1}{e}$
is the unique solution of (\ref{eq:4.2}) with the initial data
$a_{0}$, $\dot{a}_{0}$. Now since $a_{0}$, $\dot{a}_{0}$ satisfy
the initial constraint (\ref{eq:4.9}), The Hamiltonian constraint
(\ref{eq:4.1}) is satisfied on the whole existence domain $[0, T]$
of $a$. This ends the proof of theorem 4.1. $\blacksquare$.\\

\section{Local Existence for the Coupled Einstein-Boltzmann System}
\subsection{Equations and Functional Framework}
 Given the equivalence of the initial value problems
(\ref{eq:4.1})-(\ref{eq:4.2})-(\ref{eq:4.8}) with constraint
(\ref{eq:4.9})-(\ref{eq:4.10}) and
 (\ref{eq:4.26})-(\ref{eq:4.27})-(\ref{eq:4.28}), and following the
 study of the Boltzmann equation , paragraph 3, by the
 characteristic method that leads to (\ref{eq:3.14}), we study the
 initial value problem for the first order system:
 \begin{equation}\label{eq:5.1}
\frac{d f}{d t} = \frac{1}{p^{0}} Q(f, f)
\end{equation}
\begin{equation}\label{eq:5.2}
\dot{e} = - \frac{\theta}{3}\,e
\end{equation}
\begin{equation}\label{eq:5.3}
\dot{\theta} = - \frac{\theta^{2}}{3} - 4\pi(\rho + 3P)
\end{equation}
with, in (\ref{eq:5.3}) $\rho = \rho(e, f)$ and $P = P(e, f)$
defined by (\ref{eq:4.21}) and (\ref{eq:4.22}). The initial data
at $t = 0$ are denoted $ f_{0} \,; \quad e_{0}  \,; \quad
\theta_{0}$ i.e.
\begin{equation}\label{eq:5.4}
f(0, \bar{p}) = f_{0}(\bar{p})\,; \quad e(0) = e_{0}\,; \quad
\theta(0) = \theta_{0}
\end{equation}
where we take
\begin{equation}\label{eq:5.5}
f_{0} \in L_{2}^{1}(\mathbb{R}^{3})\,; \quad f_{0} \geq 0 \quad
a.e; \, e_{0} = \frac{1}{a_{0}} \,; \quad \theta_{0} =
3\frac{\dot{a}_{0}}{a_{0}}
\end{equation}
With, following (\ref{eq:4.9})-(\ref{eq:4.10}), $a_{0}$,
$\dot{a}_{0}$, $f_{0}$ subject to the constraints:
\begin{equation}\label{eq:5.6}
 a_{0} \geq \frac{3}{2}, \qquad \frac{\dot{a}_{0}}{a_{0}} =
\sqrt{\frac{8\pi}{3}\int_{\mathbb{R}^{3}}a_{0}^{3}\sqrt{1 +
a_{0}^{2}|p|^{2}}f_{0}(\bar{p})d\bar{p} + \frac{\Lambda}{3}}
\end{equation}
We are going to prove the existence of the solution $(f, e,
\theta)$ of the above initial value problem, on an interval $[0,
l]$, $l > 0$. Given the study in paragraphs 3 and 4, the
functional framework will be the Banach space $E =
L_{2}^{1}(\mathbb{R}^{3})\times \mathbb{R} \times \mathbb{R}$,
endowed with the norm
\begin{equation*}
\|(f , e , \theta)\|_{E} = \| f \| + |e| + |\theta|
\end{equation*}
where $\|.\|$ is defined by (\ref{eq:2.4}) and $|.|$ is the
absolute value in $\mathbb{R}$.
(\ref{eq:5.1})-(\ref{eq:5.2})-(\ref{eq:5.3})-(\ref{eq:5.4}) can be
written in the following standard form of first  order
differential systems:
\begin{equation}\label{eq:5.7}
  \begin{cases}
    \frac{dX}{dt} &= F(t , X) \\
    X(0) & = X_{0}
  \end{cases}
\qquad with \qquad
\begin{cases}
    X = (f , e , \theta)\, \in \, E \\
    X(0)  = (f_{0} , e_{0} , \theta_{0})
  \end{cases}
\end{equation}
We will prove the local existence theorem by applying the standard
theorem for such systems in a Banach space, and that requires that
F be continuous with respect to t and locally Lipschitzian with
respect to $(f, e, \theta)$ in the Banach space norm. In our case,
the change of variable $e = \frac{1}{a}$ shows that $p^{0} =
\sqrt{1 + a^{2}|\bar{p}|^{2}}$ and the collision kernel A given by
(\ref{eq:3.4}) depends on $e$, So that, in
(\ref{eq:5.1})-(\ref{eq:5.2})-(\ref{eq:5.3}), Q, $\rho$, P depend
only on f and e, and this implies that in (\ref{eq:5.7}), F does
not depend explicitly on t, but only on f, e, $\theta$
\subsection{The local Existence Theorem}
\begin{proposition}
There exist an interval $[0, l]$, $l > 0$ such that, the initial
value problem for the system
(\ref{eq:5.1})-(\ref{eq:5.2})-(\ref{eq:5.3}) with initial data \\
$(f_{0}, e_{0}, \theta_{0}) \in L_{2}^{1}(\mathbb{R}^{3})\times
\mathbb{R} \times \mathbb{R}$ has a unique solution $(f, e,
\theta)$ on $[0, l]$
\end{proposition}
\begin{remark}
For the proof of proposition 5.1, we will set
\begin{equation}\label{eq:5.81}
  F_{1}(e , f ,\theta) = \frac{1}{p^{0}}Q(f , f),
    \quad
    F_{2}(e , f ,\theta) = - \frac{\theta}{3}e,
    \quad
    F_{3}(e , f ,\theta) = - \frac{\theta^{2}}{3} - 4 \pi (\rho +
    3P)+ \Lambda
\end{equation}
in $F_{2}$ and $F_{3}$, $(f, e, \theta) \mapsto -
\frac{\theta}{3}e$ and $(f, e, \theta) \mapsto -
\frac{\theta^{2}}{3} + \Lambda$ are $C^{\infty}$ functions , which
are then, locally Lipschitzian with respect to $(f, e, \theta )$.
The problem will be to prove that , in $F_{1}$ and $F_{3}$, $(f,
e) \mapsto \frac{1}{p^{0}}Q(f , f)$ and $(f, e) \mapsto - 4 \pi
(\rho + 3P)(f, e)$ are locally Lipschitzian with respect to (f,
e), in the $L_{2}^{1}(\mathbb{R}^{3})\times \mathbb{R}$ norm.\\
We will deal with the quantities:
\begin{equation*}
\frac{1}{p^{0}(e_{1})}Q(f_{1}, f_{1})(e_{1}) -
\frac{1}{p^{0}(e_{2})}Q(f_{2}, f_{2})(e_{2}); \quad \rho(e_{1},
f_{1}) - \rho(e_{2}, f_{2}); \quad P(e_{1}, f_{1}) - P(e_{2},
f_{2})
\end{equation*}
and this will lead us to look in detail inside the collision
operator Q. It then shows useful to begin by giving some
preliminary results.
\end{remark}
\begin{lemma}
We have for $f, f_{1}, f_{2} \in L_{2}^{1}(\mathbb{R}^{3})$,
$e_{1}, e_{2} \in ]0, \frac{3}{2}]$, and C being a constant:
\begin{equation}\label{eq:5.8}
\left\|\left(\frac{1}{p^{0}(e_{1})} -
\frac{1}{p^{0}(e_{2})}\right)Q(f , f)(e_{2})\right\| \leq
\frac{C}{e_{1}^{2}e_{2}^{5}}\|f\||e_{1} - e_{2}|
\end{equation}
\begin{equation}\label{eq:5.9}
\left\|\frac{Q(f_{1} , f_{1})(e_{2}) - Q(f_{2} ,
f_{2})(e_{2})}{p^{0}(e_{2})}\right\| \leq
\frac{C}{e_{2}^{3}}(\|f_{1}\| + \|f_{2}\|)\|f_{2} - f_{1}\|
\end{equation}
\end{lemma}
\textbf{Proof of the lemma 5.1}\\
Recall that $p^{0} = \sqrt{1 +
\frac{1}{e^{2}}|\bar{p}|^{2}}$:\\
1) we have:
\begin{equation*}
\begin{aligned}
\left|\left(\frac{1}{p^{0}(e_{1})} -
\frac{1}{p^{0}(e_{2})}\right)Q(f , f)(e_{2})\right| &=
\left|\frac{p^{0}(e_{2}) - p^{0}(e_{1})}{p^{0}(e_{1})} \times
\frac{Q(f ,
f)(e_{2})}{p^{0}(e_{2})}\right|\\
& = \left|\frac{1}{e_{2}^{2}} - \frac{1}{e_{2}^{2}}\right|\times
\frac{|\bar{p}|^{2}}{(p^{0}(e_{1}) +
p^{0}(e_{2}))p^{0}(e_{1})}\left|\frac{Q(f ,
f)(e_{2})}{p^{0}(e_{2})}\right|
\end{aligned}
\end{equation*}
But we have:
\begin{equation*}
\frac{|\bar{p}|^{2}}{(p^{0}(e_{1}) + p^{0}(e_{2}))p^{0}(e_{1})}
\leq \frac{|\bar{p}|^{2}}{(p^{0}(e_{1}))^{2}} \leq
\frac{|\bar{p}|^{2}}{|\bar{p}|^{2}}e_{1}^{2} = e_{1}^{2} < 1
\end{equation*}
So we have, since $0< e_{i} < 1$ and using the expression
(\ref{eq:2.4}) of $\|.\|$:
\begin{equation*}
\left\|\left(\frac{1}{p^{0}(e_{1})} -
\frac{1}{p^{0}(e_{2})}\right)Q(f , f)(e_{2})\right\| \leq
\frac{2}{e_{1}^{2}e_{2}^{2}}\left\|\frac{Q(f ,
f)(e_{2})}{p^{0}(e_{2})}\right\||e_{1} - e_{2}| \tag{a}
\end{equation*}
(\ref{eq:5.8}) then follows from (a), the inequality
(\ref{eq:3.26}) on Q, in which we keep f, we set $g = 0$, and with
$a = \frac{1}{e_{2}}$ in
(\ref{eq:3.27}).\\
2) (\ref{eq:5.9}) follows from (\ref{eq:3.26}) in which we set: $f
= f_{1}$, $g = f_{2}$ and $a = \frac{1}{e_{2}}$ in
(\ref{eq:3.27}). $\blacksquare$
\begin{lemma}
We have for $f \in L_{2}^{1}(\mathbb{R}^{3})$, $e_{1}, e_{2} \in
]0, \frac{3}{2}]$ and C being a constant:
\begin{equation}\label{eq:5.10}
\left\|\frac{1}{p^{0}(e_{1})}\left[Q(f , f)(e_{1}) - Q(f ,
f)(e_{2})\right]\right\| \leq
\frac{C\|f\|^{2}}{e_{1}^{5}e_{2}^{3}}|e_{1} - e_{2}|
\end{equation}
\end{lemma}
\textbf{Proof of Lemma 5.2}\\
Here we use $Q = Q^{+} - Q^{-}$, where $Q^{+}, Q^{-}$ are given by
(\ref{eq:3.2}) and (\ref{eq:3.3}), in which $a = \frac{1}{e}$. We
write:
\begin{equation*}
Q(f, f)(e_{1}) - Q(f, f)(e_{2}) = [Q^{+}(f, f)(e_{1}) - Q^{+}(f,
f)(e_{2})] + [Q^{-}(f, f)(e_{2}) - Q^{-}(f, f)(e_{1})] \tag{a}
\end{equation*}
We can, write, using the expression (\ref{eq:3.2}) of $Q^{+}$
\begin{equation*}
 [Q^{+}(f, f)(e_{1}) - Q^{+}(f, f)(e_{2})] =
 \int_{\mathbb{R}^{3}}\int_{S^{2}}B(e_{1}, e_{2}, \bar{p}, \bar{q}, \bar{p}',
 \bar{q}')f(\bar{p}')f(\bar{q}')d\bar{q}d\omega \tag{b}
\end{equation*}
where
\begin{equation*}
B= B(e_{1}, e_{2}, \bar{p}, \bar{q}, \bar{p}',
 \bar{q}') = \frac{1}{e_{1}^{3}q^{0}(e_{1})}[A(e_{1}) - A(e_{2})]
 + \left(\frac{1}{e_{1}^{3}q^{0}(e_{1})} -
 \frac{1}{e_{2}^{3}q^{0}(e_{2})}\right)A(e_{2}) \tag{c}
\end{equation*}
in which $A(e_{i})$ stands in fact for $A(e_{i}, \bar{p}, \bar{q},
\bar{p}', \bar{q}')$, $i = 1, 2$. Assumption (\ref{eq:3.7}) on the
collision kernel A, gives:
\begin{equation*}
\left| \frac{1}{e_{1}^{3}q^{0}(e_{1})}(A(e_{1}) - A(e_{2}))
\right| \leq \frac{\gamma}{e_{1}^{3}q^{0}(e_{1})}|e_{1} - e_{2}|
\tag{d}
\end{equation*}
Now we can write,  using $q^{0}(e_{i}) = \sqrt{1 +
\frac{1}{e_{i}^{2}}|\bar{q}|^{2}} > \frac{|\bar{q}|}{e_{i}}$ and
$0 < e_{i} < 1$, $i = 1, 2$:
\begin{equation*}
\begin{aligned}
\left| \frac{1}{e_{1}^{3}q^{0}(e_{1})} -
\frac{1}{e_{2}^{3}q^{0}(e_{2})} \right| &=  \left|
\frac{(e_{2}^{3} - e_{1}^{3}) q^{0}(e_{2}) +
e_{1}^{3}[q^{0}(e_{2}) -
q^{0}(e_{1})]}{(e_{1}e_{2})^{3}q^{0}(e_{1})q^{0}(e_{2})}\right|\\
&\leq \frac{1}{q^{0}(e_{1})}\left[\frac{3|e_{1} -
e_{2}|}{(e_{1}e_{2})^{3}} + \frac{|\frac{1}{e_{2}^{2}} -
\frac{1}{e_{1}^{2}}||\bar{q}|^{2}}{e_{1}^{3}e_{2}^{3}(q^{0}(e_{2}))^{2}}\right]
\end{aligned}
\end{equation*}
Hence:
\begin{equation*}
\left| \frac{1}{e_{1}^{3}q^{0}(e_{1})} -
\frac{1}{e_{2}^{3}q^{0}(e_{2})} \right| \leq
\frac{1}{e_{1}^{3}q^{0}(e_{1})}\left( \frac{3}{e_{2}^{3}} +
\frac{2}{e_{1}^{2}e_{2}^{3}}\right)|e_{1} - e_{2}| \tag{e}
\end{equation*}
(c) then gives, using (d), (e), assumption (\ref{eq:3.4}) on A,
$(|A| \leq C_{1})$, and $0< e_{i} < 1$:
\begin{equation*}
|B| \leq
\left(\frac{1}{e_{1}^{3}q^{0}(e_{1})}\right)\frac{C}{e_{1}^{2}e_{2}^{3}}|e_{1}
- e_{2}| \tag{f}
\end{equation*}
with $C = \gamma + 5C_{1}$. We then deduce from (b) and (f), using
expression \ref{eq:2.4} of $\|.\|$
\begin{equation*}
\|\frac{[Q^{+}(f_{1} , f_{1})(e_{1}) - Q^{+}(f_{1} , f_{1})(e_{2})
]}{p^{0}(e_{1})}\| \leq \frac{C|e_{1} -
e_{2}|}{e_{1}^{2}e_{2}^{3}}\int_{\mathbb{R}^{3}}\int_{\mathbb{R}^{3}}\int_{S^{2}}\frac{\sqrt{1
+
|\bar{p}|^{2}}|f(\bar{p}')||f(\bar{q}')|}{e_{1}^{3}p^{0}(e_{1})q^{0}(e_{1})}d\bar{p}d\bar{q}d\omega
\tag{g}
\end{equation*}
We compute this integral the same way as in the proof of
(\ref{eq:3.24}) in lemma 3.1, using the change of variable
$(\bar{p}, \bar{q}) \mapsto (\bar{p}', \bar{q}')$ defined by
(\ref{eq:3.11}) and this leads, using (\ref{eq:3.27}) to
\begin{equation*}
\|\frac{[Q^{+}(f_{1} , f_{1})(e_{1}) - Q^{+}(f_{1} , f_{1})(e_{2})
]}{p^{0}(e_{1})}\|  \leq \frac{C\|f\|^{2}|e_{1} -
e_{2}|}{e_{1}^{5}e_{2}^{3}} \tag{h}
\end{equation*}
Next, we proceed the same way for the second term $Q^{-}(f,
f)(e_{2}) - Q^{-}(f, f)(e_{1})$ of (a), using this time the
expression (\ref{eq:3.3}) of $Q^{-}$. The only difference is that,
in integrals (b) and (g) $f(\bar{p}')f(\bar{q}')$ is replaced by
$f(\bar{p})f(\bar{q})$. A direct computation without change of
variable, then leads to the same estimation (h) where $Q^{+}$ is
replaced by $Q^{-}$, and lemma 5.2 follows. $\blacksquare$\\
\begin{lemma}
We have for $e_{1}, e_{2} \in ]0, \frac{2}{3}]$, $f_{1}, f_{2} \in
L^{1}_{2}(\mathbb{R}^{3})$ and $C > 0$ being a constant.
\begin{equation}\label{eq:5.11}
\left\|\frac{Q(f_{1}, f_{1})(e_{1})}{p^{0}(e_{1})} -
\frac{Q(f_{2}, f_{2})(e_{2})}{p^{0}(e_{2})}\right\| \leq
\frac{C(\|f_{1}\| + \|f_{1}\|^{2} + \|f_{2}\|)(|e_{1} - e_{2}| +
\|f_{1} - f_{2}\|)}{e_{1}^{5}e_{2}^{5}}
\end{equation}
\end{lemma}
\textbf{Proof of the lemma 5.3}\\
We can write:
\begin{equation*}
\begin{aligned}
\frac{Q(f_{1}, f_{1})(e_{1})}{p^{0}(e_{1})} - \frac{Q(f_{2},
f_{2})(e_{2})}{p^{0}(e_{2})} &= \frac{Q(f_{1}, f_{1})(e_{1}) -
Q(f_{1}, f_{1})(e_{2}) }{p^{0}(e_{1})} +
\left(\frac{1}{p^{0}(e_{1})} -
\frac{1}{p^{0}(e_{2})}\right) Q(f_{1}, f_{1})(e_{2})\\
&+ \frac{1}{p^{0}(e_{2})}[Q(f_{1}, f_{1})(e_{2}) - Q(f_{2},
f_{2})(e_{2})]
\end{aligned}
\end{equation*}
Then apply (\ref{eq:5.10}) with $f = f_{1}$ to the first term,
(\ref{eq:5.8}) with $f = f_{1}$ to the second term, and
(\ref{eq:5.9}) to the third term to obtain (\ref{eq:5.11}), by
addition of the inequalities and using $0 < e_{i} < 1$:\\
We now consider $F_{3}$ in (\ref{eq:5.81}). We can write since, by
(\ref{eq:4.21}) and (\ref{eq:4.22}), $\rho$ and P are linear in f:
\begin{equation}\label{eq:5.12}
\rho(e_{1} , f_{1}) - \rho(e_{2}, f_{2}) = \rho(e_{1} , f_{1} -
 f_{2}) + [\rho(e_{1} , f_{2}) - \rho(e_{2} , f_{2})]
\end{equation}
\begin{equation}\label{eq:5.13}
P(e_{1} , f_{1}) - P(e_{2} , f_{2}) = P(e_{1} , f_{1} -  f_{2}) +
[P(e_{1} , f_{2}) - P(e_{2} , f_{2})]
\end{equation}
(\ref{eq:4.21}) gives, using $\sqrt{1 +
\frac{|\bar{p}|^{2}}{e^{2}}} \leq \frac{1}{e}\sqrt{1 +
|\bar{p}|^{2}}$:
\begin{equation}\label{eq:5.14}
|\rho(e_{1}, f_{1} - f_{2})| \leq \frac{1}{e_{2}^{4}}\|f_{1} -
f_{2}\|
\end{equation}
(\ref{eq:5.14}) gives, using (\ref{eq:4.23}) (i.e $P \leq \rho$):
\begin{equation}\label{eq:5.15}
|P(e_{1} , f_{1} -  f_{2})| \leq \frac{1}{e_{2}^{4}}\|f_{1} -
f_{2}\|
\end{equation}
We then deduce from (\ref{eq:5.12}) and (\ref{eq:5.13}), using
(\ref{eq:4.29}), (\ref{eq:4.30}) in which we set $f = f_{2}$,
(\ref{eq:5.14}), (\ref{eq:5.15}) and $0< e_{i} < 1$, i = 1 , 2,
that
\begin{equation}\label{eq:5.16}
|\rho(e_{1} , f_{1}) - \rho(e_{2} , f_{2})| \leq \frac{C(\|f_{1}\|
+ 1)}{e_{1}^{4}e_{2}^{6}}(|e_{2} - e_{1}| + \|f_{1} - f_{2}\|)
\end{equation}
\begin{equation}\label{eq:5.17}
|P(e_{1} , f_{1}) - P(e_{2} , f_{2})| \leq \frac{C(\|f_{1}\| +
1)}{e_{1}^{4}e_{2}^{7}}(|e_{2} - e_{1}| + \|f_{1} - f_{2}\|)
\end{equation}
Now let $f^{1} \in L^{1}_{2}(\mathbb{R}^{3})$ and $e^{1} \in ]0,
\frac{3}{2}]$; take $e_{1}, e_{2} \in ]\frac{e^{1}}{2},
\frac{e^{1} + \frac{3}{2}}{2}[$ and \\
$f_{1}, f_{2} \in B(f^{1}, 1)
:= \{f \in L_{2}^{1}(\mathbb{R}^{3}) \|f - f^{1}\ < 1|\}$\\
Then we have , $\frac{1}{e_{1}}, \frac{1}{e_{2}} <
\frac{2}{e^{1}}$ and $\|f_{1}\|, \|f_{2}\|
< \|f^{1}\| + 1$.\\
We then deduce from (\ref{eq:5.11}), (\ref{eq:5.16})  and
(\ref{eq:5.17}) that:
\begin{equation*}
\left\|\frac{Q(f_{1} , f_{1})(e_{1})}{p^{0}(e_{1})} -
\frac{Q(f_{2} , f_{2})(e_{2})}{p^{0}(e_{2})}\right\| \leq
K(\|f_{1} - f_{2}\| + |e_{1} - e_{2}|)
\end{equation*}
\begin{equation*}
\left|\rho(e_{1}, f_{1}) - \rho(e_{2}, f_{2})\right| \leq
K(\|f_{1} - f_{2}\| + |e_{1} - e_{2}|)
\end{equation*}
\begin{equation*}
\left|P(e_{1}, f_{1}) - P(e_{2}, f_{2})\right| \leq K(\|f_{1} -
f_{2}\| + |e_{1} - e_{2}|)
\end{equation*}
where $K = K(e^{1}, f^{1})$ is a constant depending only on
$e^{1}$ and $f^{1}$. If we add this to the fact that $(f, e,
\theta) \mapsto -\frac{\theta}{3}e$ and $(f, e, \theta) \mapsto
-\frac{\theta^{2}}{3} + \Lambda$ are Lipschitzian with respect to
the norm of $E = L_{2}^{1}(\mathbb{R}^{3}) \times \mathbb{R}\times
\mathbb{R}$, we can conclude that $F = (F_{1}, F_{2}, F_{3})$
defined by (\ref{eq:5.8}) is locally Lipschitzian in $(f, e,
\theta)$ with respect to the norm of E. Proposition 5.1 then
follows from the standard theorem on first order differential
system for functions with values in a Banach space. Notice the $e,
\theta$ are continuous and $f \in
C[0, l;L_{2}^{1}(\mathbb{R}^{3})]$ $\blacksquare$\\
From proposition 5.1, we deduce the following theorem:
\begin{theorem}
Let $f_{0} \in L_{2}^{1}(\mathbb{R}^{3})$, $f_{0} \geq 0$ a.e,
$a_{0} \geq \frac{3}{2}$ and let a strictly positive cosmological
constant $\Lambda$, be given. Define $\dot{a}_{0}$ by the relation
(\ref{eq:5.6}). \\
Let $r > \|f_{0}\|$ . Then, there exist a number $l > 0$ such that
the initial value problem for the coupled Einstein-Boltzmann
system (\ref{eq:2.13})-(\ref{eq:4.1})-(\ref{eq:4.2}) with the the
initial data $(f_{0}, a_{0}, \dot{a}_{0})$, has an unique solution
(f, a) on $[0, l]$. The solution (a, f) has the following
properties:
\begin{itemize}
    \item[(i)] a is an increasing function.
    \item[(ii)]
\begin{equation}\label{eq:5.18}
    f \in C[[0, l], X_{r}]
\end{equation}
    \item[(iii)]
\begin{equation}\label{eq:5.19}
     \||f\|| \leq \|f_{0}\|
\end{equation}
\end{itemize}
\end{theorem}
\textbf{Proof of theorem 5.1}\\
Choose in proposition 5.1, $f_{0}, e_{0}$ and $\theta_{0}$ as in
(\ref{eq:5.5}). Let $(e, f, \theta)$ be the unique solution of
that initial value problem defined on the interval $[0, l]$, $l >
0$, whose existence is proved by that proposition. Set $a =
\frac{1}{e}$, (\ref{eq:5.2}) then implies $\theta = 3
\frac{\dot{a}}{a}$. We know, by the characteristic method that the
Boltzmann equation (\ref{eq:2.13}) is equivalent to
(\ref{eq:5.1}). We also know that, in the framework described in
$\S 4.4$, and in which we solve the Einstein equation in the case
$\Lambda > 0$, the system (\ref{eq:4.1})-(\ref{eq:4.2}) in a, and
the system (\ref{eq:5.2})-(\ref{eq:5.3}) in $(e, \theta)$ are
equivalent. Since the relation (\ref{eq:5.6}) which implies
(\ref{eq:4.9}), is satisfied, we know that the Hamiltonian
constraint (\ref{eq:4.1}) is satisfied on $[0, l]$. We can
conclude that, in the framework fixed by choosing (\ref{eq:4.16})
in the case $\Lambda > 0$ , the unique solution $(f, e, \theta)$
on $[0, l]$, of the initial value problem
(\ref{eq:5.1})-(\ref{eq:5.2})-(\ref{eq:5.3})-(\ref{eq:5.4}) with
$f_{0}, e_{0}, \theta_{0}$given by (\ref{eq:5.5}) and satisfying
(\ref{eq:5.6}), give the unique solution $(f, a = \frac{1}{e})$ on
$[0, l]$, of the initial value problem
(\ref{eq:2.13})-(\ref{eq:4.1})-(\ref{eq:4.2}) with the initial
data
$(f_{0}, a_{0}, \dot{a}_{0})$ satisfying (\ref{eq:5.6}).\\
Concerning the properties of the solution (f, a) on $[0, l[$:
 (\ref{eq:4.16}) shows that $\dot{a} > 0$; $t \mapsto a(t)$ then satisfies the
point i) and a is bounded from below, since $a(t) \geq a_{0} \geq
\frac{3}{2}$. a then satisfies the hypotheses  of theorem 3.1 in
which we set $t_{0} = 0, T = l$. Note that $C[[0, l]; X_{r}]
\subset C[[0, l]; L_{2}^{1}(\mathbb{R}^{3})]$; (\ref{eq:5.18})
follows from the uniqueness and (\ref{eq:5.19}) from
(\ref{eq:3.18}).$\blacksquare$
\section{Global Existence Theorem for the Coupled Einstein-Boltzmann System}
\subsection{The Method}
Here, we prove that the local solution obtained in $\S 5$ is, in
fact, a global solution. Let us sketch the method we adopt: Denote
$[0, T[$, where $T > 0$, the \textbf{maximal} existence domain of
the solution  of (\ref{eq:5.1})-(\ref{eq:5.2})-(\ref{eq:5.3}),
with initial data $(f_{0}, e_{0}, \theta_{0})$ defined by
(\ref{eq:5.4})-(\ref{eq:5.5})-(\ref{eq:5.6}); here we denote this
solution $(\tilde{f}, \tilde{e}, \tilde{\theta})$, in order words
we have, on $[0, T[$
\begin{equation}\label{eq:6.1}
\dot{\tilde{f}} = \frac{1}{p^{0}(\tilde{e})}Q(\tilde{f},
\tilde{f})
\end{equation}
\begin{equation}\label{eq:6.2}
\dot{\tilde{e}} = - \frac{\tilde{\theta}}{3}\tilde{e}
\end{equation}
\begin{equation}\label{eq:6.3}
\dot{\tilde{\theta}} = - \frac{\tilde{\theta}^{2}}{3} - 4 \pi
(\tilde{\rho} + 3\tilde{P}) + \Lambda
\end{equation}
\begin{equation}\label{eq:6.4}
\tilde{f}(0) = f_{0} \in X_{r}, \qquad \tilde{e}(0) = e_{0} =
\frac{1}{a_{0}}, \qquad \tilde{\theta}(0) =
3\frac{\dot{a}_{0}}{a_{0}}
\end{equation}
with in (\ref{eq:6.3}) $\tilde{\rho} = \rho(\tilde{f},
\tilde{e})$: $\tilde{P} = P(\tilde{f}, \tilde{e})$ and in
(\ref{eq:6.4}) $f_{0}$, $a_{0}$, $\theta_{0}$ subject to the constraints (\ref{eq:5.6}). $r > 0$ is given such that $r > \|f_{0}\|$.\\
If $T = + \infty$, the problem is solved. We are going to show
that, if we suppose that $T < + \infty$, then the solution
$(\tilde{f}, \tilde{e}, \tilde{\theta})$ can be extended beyond T,
which contradicts the maximality of T. Suppose $0< T < + \infty$
and let $t_{0} \in [0, T[$. We will show that, there exists a
strictly positive number $\delta > 0$, \textbf{independent of
$t_{0}$}, such that the following system in $(e, \theta)$ on
$[t_{0}, t_{0} + \delta]$, in which $\tilde{a} =
\frac{1}{\tilde{e}}$:
\begin{equation}\label{eq:6.5}
\dot{f} = \frac{1}{p^{0}(e)}Q(f, f)
\end{equation}
\begin{equation}\label{eq:6.6}
\dot{e} = - \frac{\theta}{3}\,e
\end{equation}
\begin{equation}\label{eq:6.7}
\dot{\theta} = - \frac{\theta^{2}}{3} - 4\pi (\rho + 3 P) +
\Lambda
\end{equation}
\begin{equation}\label{eq:6.8}
f(t_{0}) = \tilde{f}(t_{0}),\quad e(t_{0})= \tilde{e}(t_{0}),
\quad \theta(t_{0}) = \tilde{\theta}(t_{0}) = 3
\frac{\dot{\tilde{a}}(t_{0})}{\tilde{a}(t_{0})}
\end{equation}
has a solution $(f, e, \theta)$ on $[t_{0}, t_{0} + \delta]$.
Then, by taking $t_{0}$ sufficiently close to T, for example, to
such that $0 < T - t_{0} < \frac{\delta}{2}$, hence $T < t_{0} +
\frac{\delta}{2}$, we can extend the solution $(\tilde{f},
\tilde{e}, \tilde{\theta})$ to $[0, t_{0} + \frac{\delta}{2}[$
that contains strictly $[0, T[$, and this will contradicts the
maximality of T.We need some preliminaries results.\\
In what follows, we suppose $0 < T < + \infty$ and $t_{0} \in [0,
T[$.

\subsection{The Functional Framework}
In all what follows, $C_{2}$, $C_{3}$ , $D_{0}$ are the absolute
constant defined by (\ref{eq:4.285}) and (\ref{eq:4.37}). We set,
for $\delta > 0$:
\begin{equation*}
E_{t_{0}}^{\delta} = \left\{e \in C[t_{0} , t_{0} +\delta ] ,
\frac{1}{C_{2}}e^{-C_{3}(t_{0} + t + 1)^{2}} \leq e(t_{0} + t)
\leq \frac{2}{3} \quad, \forall t \in [0 , \delta[ \right\}
\end{equation*}
\begin{equation*}
F_{t_{0}}^{\delta} = \left\{\theta \in C[t_{0} , t_{0} +\delta ] ,
\, \theta \,  , \, \sqrt{3\Lambda} \leq \theta(t_{0} + t) \leq
D_{0}\, \quad \forall t \in [0 , \delta[\right\}
\end{equation*}
where $C[t_{0} , t_{0} +\delta ]$ is the space of continuous (and
hence bounded) functions on $[t_{0} , t_{0} +\delta] $. One
verifies easily that $E_{t_{0}}^{\delta}$ and $F_{t_{0}}^{\delta}$
are complete metric subspaces of the Banach space $(C[t_{0} ,
t_{0} +\delta ], \|.\|_{\infty})$ where $\|u\|_{\infty} =
\underset{t \in [t_{0} , t_{0} +\delta ]}{Sup}|u(t)|$
\subsection{The Global Existence Theorem}
\begin{proposition}
There exists a strictly positive real number  $\delta > 0$
depending only on the absolute constants $a_{0}$, $\Lambda$ , $r$
 and $T$ such that the initial value
problem
(\ref{eq:6.5})-(\ref{eq:6.6})-(\ref{eq:6.7})-(\ref{eq:6.8}) has a
solution $(f , e = \frac{1}{a} , \theta)\, \in C[[t_{0} , t_{0} +
\delta] ; X_{r}] \times E_{t_{0}}^{\delta} \times
F_{t_{0}}^{\delta}$
\end{proposition}
\textbf{Proof}\\
It will be enough, if we look for $\delta$ such that $0 < \delta <
1$. By theorem 3.1, we know that if we fix $\bar{e} \in
E_{t_{0}}^{\delta}$ and if we set $\bar{a} = \frac{1}{\bar{e}}$,
then (\ref{eq:6.5}) has an unique solution $f \in C[[t_{0} , t_{0}
+ \delta]  ; X_{r}]$, such that, $f(t_{0}) = \tilde{f}(t_{0})$,
and, by (\ref{eq:3.18}) and (\ref{eq:5.19}):
\begin{equation}\label{eq:6.18}
\|f(t)\| \leq \|\tilde{f}(t_{0})\| \leq \|f_{0}\| \leq r
\end{equation}
Next, by proposition 4.6 in which we set $\Xi = \tilde{e}, \Theta
= \tilde{\theta}$, we know that if $\bar{f}$ is given in $C[[t_{0}
, t_{0} + \delta] ; X_{r}]$, then (\ref{eq:6.6})-(\ref{eq:6.7})
has an unique solution $(e, \theta)$ on $[t_{0} , t_{0} + \delta]$
such that $e(t_{0}) = \frac{1}{\tilde{a}(t_{0})}$; $\theta(t_{0})
= 3 \frac{\dot{\tilde{a}}(t_{0})}{\tilde{a}(t_{0})}$. Now
(\ref{eq:4.38}) and (\ref{eq:4.39}) show that $(e = \frac{1}{a} ,
\theta) \in E_{t_{0}}^{\delta} \times F_{t_{0}}^{\delta}$. This
allows us to define the application :
\begin{eqnarray}\label{eq:6.19}
  G : C[[t_{0} , t_{0} + \delta]  ;
X_{r}] \times E_{t_{0}}^{\delta} &\rightarrow& C[[t_{0} , t_{0} +
\delta]  ;
X_{r}] \times (E_{t_{0}}^{\delta} \times F_{t_{0}}^{\delta}) \\
(\bar{f} , \bar{e}) &\mapsto& G(\bar{f}, \bar{e}) = [f , (e,
\theta)]
\end{eqnarray}
We are going to show that we can find $\delta > 0$ such that G
defined by (\ref{eq:6.19}) induces a contracting map of the
complete metric space $C[[t_{0} , t_{0} + \delta]  ; X_{r}] \times
E_{t_{0}}^{\delta}$ into itself, that will hence, have an unique
fixed point $(f , e)$ ; this will allow us to find  $\theta$ such
that $(f, e , \theta)$ be the unique solution of
(\ref{eq:6.5})-(\ref{eq:6.6})-(\ref{eq:6.7})-(\ref{eq:6.8}) in
$C[[t_{0} , t_{0} + \delta] ; X_{r}] \times (E_{t_{0}}^{\delta}
\times F_{t_{0}}^{\delta})$. So if we set in (\ref{eq:6.5}) $e =
\bar{e} \in E_{t_{0}}^{\delta}$, in (\ref{eq:6.7}) $\rho =
\bar{\rho} = \rho(e, \bar{f})$, $P = \bar{P} = P(e, \bar{f})$
where $\bar{f} \in C[[t_{0} , t_{0} + \delta]  ; X_{r}]$, we have
a solution $(f, e , \theta)$ of that system, or, equivalently, a
solution $(f, e , \theta)$ of the following integral system:
\begin{equation}\label{eq:6.20}
  f(t_{0} + t) = \tilde{f}(t_{0}) + \int_{t_{0}}^{t_{0} + t}\frac{1}{p^{0}(\bar{e})}Q(f , f)(\bar{e})(s)ds
\end{equation}
\begin{equation}\label{eq:6.21}
 e(t_{0} + t)  = \tilde{e}(t_{0}) + \int_{t_{0}}^{t_{0} - t}\frac{\theta(s)e(s)}{3}ds
\end{equation}
\begin{equation}\label{eq:6.22}
 \theta(t_{0} + t) = \tilde{\theta}(t_{0}) + \int_{t_{0}}^{t_{0} +
 t}[- \frac{\theta^{2}}{3} - 4\pi(\bar{\rho} + 3 \bar{P}) +
 \Lambda](s)ds
\end{equation}
 $t \in [0 , \delta[$\\
Let $\bar{e}_{1} , \bar{e}_{2} \, \in \, E_{t_{0}}^{\delta}$ ;
$\bar{f}_{1} , \bar{f}_{2} \, \in \,C[[t_{0} , t_{0} + \delta]  ;
X_{r}]$. To $\bar{e}_{i}$ (resp $\bar{f}_{i}$), i = 1, 2,
corresponds by G, the solution $f_{i}$ (resp $(e_{i},
\theta_{i})$) of (\ref{eq:6.20}), [resp
(\ref{eq:6.21})-(\ref{eq:6.22})]. Writing each equation for i = 1,
i = 2 and subtracting yields:
\begin{equation}\label{eq:6.23}
\|f_{1} - f_{2}\| \leq \int_{t_{0}}^{t_{0} +
t}\left\|\frac{1}{p^{0}(\bar{e}_{1})}Q(f_{1} , f_{1})(\bar{e}_{1})
- \frac{1}{p^{0}(\bar{e}_{2})}Q(f_{2} ,
f_{2})(\bar{e}_{2})\right\|ds
\end{equation}
\begin{equation}\label{eq:6.24}
|e_{1} - e_{2}| \leq  \int_{t_{0}}^{t_{0} +
t}\left|\frac{\theta_{1}e_{1}}{3}  -
\frac{\theta_{2}e_{2}}{3}\right|ds
\end{equation}
\begin{equation}\label{eq:6.25}
|\theta_{1} - \theta_{2}| \leq \int_{t_{0}}^{t_{0} +
 t}[\frac{1}{3}|\theta_{1}^{2} - \theta_{2}^{2}| + 12 \pi(|\bar{\rho_{1}} - \bar{\rho_{2}}| + |\bar{P_{1}} -
 \bar{P_{2}}|)]ds
\end{equation}
We are going to bound the r.h.s of (\ref{eq:6.23}) and
(\ref{eq:6.25}), using:  (\ref{eq:5.11}) in which we set $e_{1} =
\bar{e}_{1}$, $e_{2} = \bar{e}_{2}$ and
(\ref{eq:5.16})-(\ref{eq:5.17}), in which we set $f_{1} =
\bar{f}_{1}$, $f_{2} = \bar{f}_{2}$. Since $\bar{e}_{1},
\bar{e}_{2}, e_{1}, e_{2} \in E_{t_{0}}^{\delta}$,
(\ref{eq:4.38}), shows that: $\frac{1}{e_{i}},
\frac{1}{\bar{e}_{i}} \leq C_{2}e^{C_{3}(T + 2)^{2}}$, since
$t_{0} < T$, $t \leq \delta < 1$. So we deduce from
(\ref{eq:6.23}), (\ref{eq:6.24}) and (\ref{eq:6.25}), using
(\ref{eq:5.11}), (\ref{eq:5.16}), (\ref{eq:5.17}),
$\||\bar{f}_{i}\|| < r$ , $0 \leq t \leq \delta$, the definition
of $E_{t_{0}}^{\delta}$, $F_{t_{0}}^{\delta}$ ($|\theta_{i}| \leq
D_{0}(a_{0}, r, \Lambda, T)$), see (\ref{eq:4.37}), the definition
(\ref{eq:2.16}) of the norm $\||.\||$ of function over $[t_{0},
t_{0} + \delta]$, that:
\begin{equation}\label{eq:6.26}
\||f_{1} - f_{2}\|| \leq \delta M_{1}(\|\bar{e}_{1} -
\bar{e}_{2}\|_{\infty} + \||f_{1}  - f_{2}\||)
\end{equation}
\begin{equation}\label{eq:6.27}
\|e_{1} - e_{2}\|_{\infty} \leq \delta M_{2}(\|e_{1} -
e_{2}\|_{\infty} + \|\theta_{1} - \theta_{2}\|_{\infty})
\end{equation}
\begin{equation}\label{eq:6.28}
\|\theta_{1} - \theta_{2}\|_{\infty} \leq \delta M_{3} (\|e_{1} -
e_{2}\|_{\infty} + \|\theta_{1} - \theta_{2}\|_{\infty}  +
\||\bar{f}_{1} - \bar{f}_{2}\||)
\end{equation}
where $M_{1}$, $M_{2}$, $M_{3}$ are constants depending only on
$a_{0}, \Lambda, r$ and T. We have by addition of (\ref{eq:6.27})
and (\ref{eq:6.28}):
\begin{equation}\label{eq:6.30}
\|e_{1} - e_{2}\|_{\infty} + \|\theta_{1} - \theta_{2}\|_{\infty})
\leq 2\delta (M_{2} + M_{3})(\|e_{1} - e_{2}\|_{\infty} +
\|\theta_{1} - \theta_{2}\|_{\infty} + \||\bar{f}_{1} -
\bar{f}_{2}\||)
\end{equation}
Then, if we choose $\delta$ such that:
\begin{equation}\label{eq:6.30}
\delta  = Inf\left[1 , \frac{1}{8(M_{1} + M_{2} + M_{3})}\right]
\end{equation}
(\ref{eq:6.30}) implies that : $\delta M_{1} < \frac{1}{4}$;
$2\delta (M_{2} + M_{3}) < \frac{1}{4}$ and (\ref{eq:6.26}),
(\ref{eq:6.29}) give:
\begin{equation*}
\begin{cases}
\||f_{1} - f_{2}\|| &\leq \frac{1}{3}\|\bar{e}_{1} -
\bar{e}_{2}\|_{\infty}  \\
\\
\|e_{1} - e_{2}\|_{\infty} + \|\theta_{1} - \theta_{2}\|_{\infty}
&\leq   \frac{1}{3}\||\bar{f}_{1} - \bar{f}_{2}\||
\end{cases}
\end{equation*}
and by addition:
\begin{equation*}
\||f_{1} - f_{2}\|| + \|e_{1} - e_{2}\|_{\infty} + \|\theta_{1} -
\theta_{2}\|_{\infty} \leq \frac{1}{3}(\||\bar{f}_{1} -
\bar{f}_{2}\|| + \|\bar{e}_{1} - \bar{e}_{2}\|_{\infty})
\end{equation*}
from which, we deduce:
\begin{equation}\label{eq:6.31}
\||f_{1} - f_{2}\|| + \|e_{1} - e_{2}\|_{\infty}  \leq
\frac{1}{3}(\||\bar{f}_{1} - \bar{f}_{2}\|| + \|\bar{e}_{1} -
\bar{e}_{2}\|_{\infty})
\end{equation}
(\ref{eq:6.31}) shows that the map $(\bar{f} , \bar{e}) \mapsto (f
, e)$ is a contracting map from the complete metric space
$C[[t_{0} , t_{0} + \delta] ; X_{r}] \times E_{t_{0}}^{\delta}$
into itself, for every $\delta$ satisfying (\ref{eq:6.30}), which
shows that such a $\delta$ depends only on $a_{0}, r, \Lambda $
and T. This map has an unique fixed point (f, e); since $e$ is
known, (\ref{eq:6.6}) determines $\theta$ by $\theta = -
3\frac{\dot{e}}{e}$, and $(f, e, \theta)$ is a solution of
(\ref{eq:6.5})-(\ref{eq:6.6})-(\ref{eq:6.7})-(\ref{eq:6.8}) in
$C[[t_{0} , t_{0} + \delta] ; X_{r}] \times
E_{t_{0}}^{\delta}\times F_{t_{0}}^{\delta}$. This complete the
proof of proposition 6.4
$\blacksquare$\\
We can then state:
\begin{theorem}
The initial value problem for the Einstein-Boltzmann system with a
strictly positive cosmological constant $\Lambda$ on a
Robertson-Walker space-time has a global solution (a, f) on $[0, +
\infty[$, for arbitrarily large initial data $a_{0}$ and $f_{0}
\in L_{2}^{1}(\mathbb{R}^{3})$, $f_{0} \geq 0$ a.e .
\end{theorem}
\begin{remark}
\begin{itemize}
    \item[1)]Nowhere in the proof we had to restrict the size of
    the initial data $a_{0}$, $f_{0}$, which can then be taken
    arbitrarily large.
    \item[2)] In \cite{n}, the author considered only the
    Hamiltonian constraint (\ref{eq:4.1}), in the case $\Lambda =
    0$, without studying the evolution equation (\ref{eq:4.2}) that
    is, as we saw, the main problem to solve, since the
    Hamiltonian constraint (\ref{eq:4.1}) is satisfied once it is
    the case for the initial data.
    \item[3)] We will prove in a future paper, that theorem 6.1
    extends to the case $\Lambda = 0$
    \item[4)] In the future, we will try to relax hypotheses on
    the collision kernel A.
\end{itemize}
\end{remark}
\textbf{\underline{Acknowledgement}}. The authors thank
A.D.Rendall for helpful comments and suggestions. This work was
supported by the VolkwagenStiftung, Federal Republic of Germany.

\end{document}